\documentclass[notitlepage]{article}

\usepackage{amssymb}
\usepackage{amsmath}
\usepackage[twoside]{geometry}
\usepackage[dvips]{graphicx}

\begin{document}

\date{}
\title{\textbf{Feynman Diagrams and a Combination of the Integration by
Parts (IBP) and the Integration by Fractional Expansion (IBFE) Techniques}\\
}
\author{Iv\'{a}n Gonz\'{a}lez\thanks{%
e-mail: igonzalez@fis.puc.cl} \ and M. Loewe\thanks{%
e-mail: mloewe@fis.puc.cl} \\
Facultad de F\'{\i}sica\\
Pontificia Universidad Cat\'{o}lica de Chile\\
Casilla 306, Santiago 22, Chile}
\maketitle

\begin{abstract}
In this paper, we show how to improve and extend the Integration by
Fractional Expansion technique (IBFE) by applying it to certain families of
scalar massive Feynman diagrams. The strategy is based on combining this
method together with the Integration by Parts technique (IBP). In
particular, we want to calculate certain Feynman diagrams which have a
triangle loop as a subgraph. The main idea is to use IBP in this subgraph in
order to simplify the topology of the original diagram in which it is
immersed, using then, in a second step, the IBFE technique. The result we
have obtained, after the application of both techniques, represents a
simplification in the complexity of the solution, compared with having used
only the IBFE technique.
\end{abstract}


\bigskip


\vfill\newpage

\section{Introduction}

\qquad Since the beginning of the development of quantum field theory, the
evaluation of Feynman diagrams has been one of the most important
theoretical challenges. The development of methods leading to the analytical
or numerical evaluation of loop integrals represents an important problem
both from the point of view of physics, as well as from the perspective of
advanced mathematical methods.

Concerning physics, this research is relevant for perturbative calculations
of higher orders, needed to test our present theoretical models when their
predictions are compared to experimental results from present and future
colliders and/or from other high energy physics experiments. The amount of
experimental information at present is so huge, and with such a high
precision, that the numerical calculations are almost not enough to achieve
the needed accuracy in order to compare to the experimental results.
Therefore, it is extremely important to be able to optimize the present
existing integration techniques and/or to generate new mathematical
techniques that may allow us to face up the calculation of Feynman diagrams
with a higher number of loops, or that may involve also the presence of
multiple scales.

In this perspective, we consider the Integration by Fractional Expansion
Technique (IBFE) as a method that allows the calculation of certain families
of Feynman diagrams in a simple and direct way, getting solutions for the
general case, i.e. arbitrary indices for propagators, in terms of
generalized hypergeometric functions. By indices we understand here, as
usual, the power degree of the propagators. In some cases, for some special
values of these indices, it is possible to carry out the sum of these series
in terms of conventional functions which simplifies the subsequent analysis
of the final solution.

The IBFE technique, in general aspects, is similar to the integration
technique based on the Mellin-Barnes representation of the diagram \cite%
{NUs, EBo}. Other authors have called this technique NDIM (Negative
Dimensional Integration Method) \cite{CAn, ASu1, ASu2}, which is based on
the work by Halliday and Ricotta \cite{IRi}. These authors make an
analytical continuation of the $D$ dimension, working in the frame of
dimensional regularization $\left( D=4-2\epsilon \right) $\thinspace\ into
negative values of $D$, which justifies the name of the method. However, we
prefer to use the name IBFE due to the reasons explained in \cite{IGoIBFE}.

It is possible to extend the range of validity of IBFE by considering a new
kind of diagrams that may be evaluated by means of this technique. In
particular, in this article we will show that with the help of the
Integration by Parts (IBP) method \cite{TKa}, it turns out that certain
families of diagrams have more simple solutions compared with those that
appear when only the IBFE method is used. It is important to remark that the
existence of these simple solutions allows us to increase the degree of
difficulty of the Feynman diagrams that can be handled with the combined
IBFE + IBP techniques. We could also be able to consider diagrams with a
certain particular mass distribution in the propagators.

To justify this methodology, we will concentrate on a certain class of
Feynman diagrams that involve in their geometry a one-loop triangle
subgraph, and such that we may introduce the IBP technique through the
so-called triangle identity. The main idea is to use IBP to transform the
geometry of the diagram. The elimination of certain propagators of the
original diagram may produce now a diagram which will be ideal for the
application of the IBFE method, as was pointed out in \cite{IGoIBFE}.
Basically, it turns out that triangles are transformed into bubbles under
some conditions determined by the indices of the propagators.

This work has been organized in the following way: In Sec. II, we remind the
formula associated to the triangle identity. This rule emerges when the IBP
technique is applied to a one-loop diagram with three external lines. Here
we have assumed a scalar massless theory.

In Sec. III, and in order to justify the combined use of IBFE + IBP
techniques, we will evaluate a simple diagram, the one-loop triangle diagram
with one massive propagator with mass $m$. To simplify the problem even more
, we have assumed that two of the external lines are on mass shell with $m=0$%
. In this way, the diagram will be characterized by only two invariants $%
\left\{ m^{2},s\right\} $. The above mentioned simplifications, as we will
see later, do not imply a loss with respect to the general statements which
are the goal of this section.

The validation of the combined use of IBFE + IBP will be realized through a
calculation of this diagram, using first only the IBFE technique. Later, the
same diagram is evaluated with the combined techniques IBFE + IBP. We will
show that both techniques provide equivalent solutions, both for numerical
comparisons for some specific values of the invariants of the problem, as
well as for analytical comparisons in certain special limits, ($m=0$ or $s=0$%
).

In Sec. IV, we present a more complex application. The idea is to specify
the range of validity of the combined use of the IBFE + IBP techniques. In
particular, we consider the topology associated to the radiative corrections
to the propagator with two loops and five internal lines. This diagram can
be easily evaluated by means of IBP when all of the propagators are massless
with index one. In our case, we will extend the discussion by considering
now this diagram with one massive propagator and allowing some arbitrary
indices. The analytic solutions for this problem will be given in terms of
generalized hypergeometric functions of the form \ $_{q}F_{q-1}$, whose
argument can be $\left( \dfrac{M^{2}}{p^{2}}\right) $ or $\left( \dfrac{p^{2}%
}{M^{2}}\right) ,$ according to the relevant kinematical region; $p$ denotes
the momentum that flows into the diagram and which fixes the scale, and $M$
is the mass in the massive propagator. The general solutions of Feynman
diagrams will correspond always to hypergeometric series, i.e. to natural
expansions around a zero value of the argument and with a convergence radius
$r=1$. The mathematical formalism, the relevant formulas, and applications
are presented and discussed in \cite{WBa,LSl,LGr,HEx}\ ).

\section{IBP and the triangle rule}

\qquad The well known triangle rule is obtained directly by applying the IBP
method to the massless one-loop diagram with three external lines (how his
rule is derived can be seen in the book by V. A. Smirnov \cite{VSm}). The
integral that represents the triangle diagram is the following:

\begin{equation}
\frac{1}{\left( P_{\alpha }\right) ^{2a_{4}}}\frac{1}{\left( P_{\beta
}\right) ^{2a_{5}}}\int \frac{d^{D}q}{i\pi ^{D/2}}\frac{1}{\left[ (P_{\alpha
}+q)^{2}\right] ^{a_{1}}}\frac{1}{\left[ (P_{\beta }-q)^{2}\right] ^{a_{2}}}%
\frac{1}{\left( q^{2}\right) ^{a_{3}}},
\end{equation}%
where $P_{\alpha }$ and $P_{\beta }$ identify the independent momenta of the
external lines. Then, associating each index $\left\{
a_{1},a_{2},a_{3}\right\} $\ with his own propagator, we may introduce a
graphical representation for the triangle identity once IBP has been applied
to the previous integral.

\begin{equation}
\begin{array}{rr}
\begin{minipage}{3.1cm} \includegraphics[scale=.7] {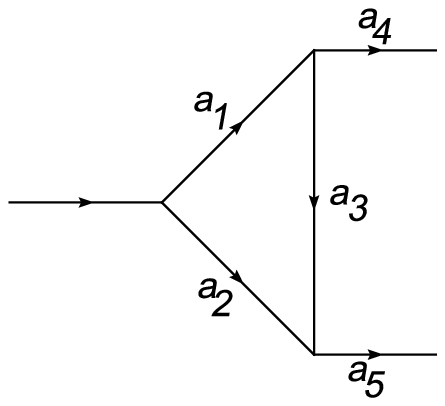}
\end{minipage}= & \dfrac{1}{\left( D-a_{1}-a_{2}-2a_{3}\right) }\left[
\;a_{1}\left(
\begin{minipage}{3.1cm} \includegraphics[scale=.7]
{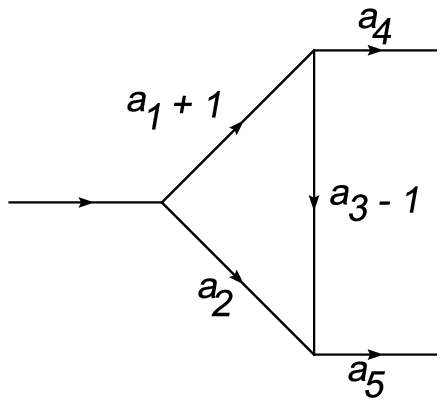} \end{minipage}-\begin{minipage}{3.1cm}
\includegraphics[scale=.7] {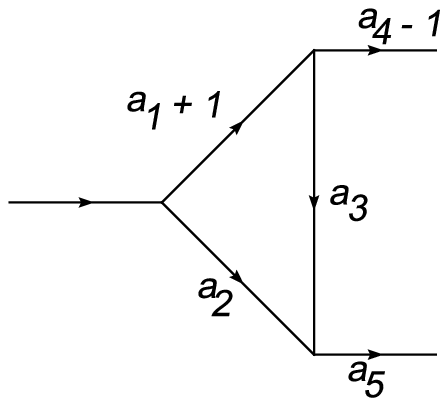} \end{minipage}\;\right) \right.
\\
&  \\
& \left. +\;a_{2}\left(
\begin{minipage}{3.1cm} \includegraphics[scale=.7]
{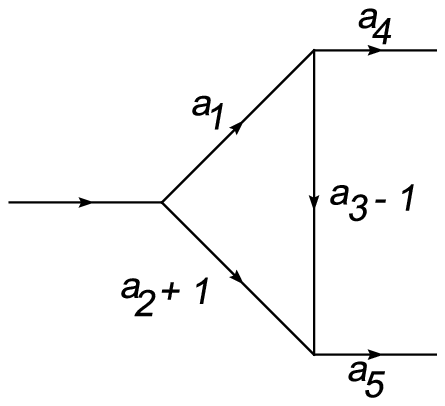} \end{minipage}-\begin{minipage}{3.1cm}
\includegraphics[scale=.7] {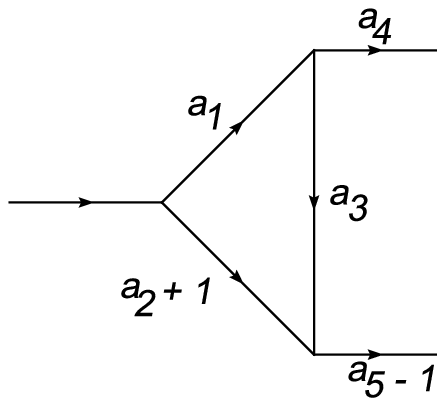} \end{minipage}\;\right) \;\right]
.%
\end{array}
\label{ibfe1}
\end{equation}%
In this expression, $D$ denotes the number of space dimensions in the
integral. The indices $a_{i}\left( i=1,2,3,4,5\right) $ correspond, as we
said, to the powers of the propagators of the corresponding internal lines
in the diagram. The indices $a_{4}$ and $a_{5}$ will be arbitrary if the
corresponding lines are internal lines in a diagram that contains this
triangle and their value will be zero when the lines are external.
Obviously, the IBP technique goes beyond this identity. However, for our
purpose this identity will be enough.

\section{Validation of the IBFE + IBP method}

In previous articles \cite{IGoIBFE, IGo},\ we have shown that the IBFE
technique is a useful method for evaluating certain families of Feynman
diagrams. In particular, we will use now this technique for handling the
following test topology: a one-loop triangle diagram where one of the
propagators has a finite mass and the two external lines are massless. The
result for this diagram will be compared with the result that emerges from
the combined IBFE and IBP techniques. Explicitly, we will expand the massive
propagator according to the IBFE prescription, then extracting out the mass
from the diagram and getting finally a series that involves only non massive
diagrams. Each of them may be reduced topologically, by means of the IBP
approach, to a series of bubble diagrams. The IBFE technique will be applied
then to each one of these bubble diagrams, obtaining the Multiregion
Expansion (MRE) \cite{IGoIBFE} of the complete diagram which allows us to
obtain the solutions in terms of hypergeometric functions of the form \ $%
_{q}F_{q-1}$.

Once we get the solution of this diagram in both ways, we will make a
analytical comparison in the limits $m=0$ and $s=0$.

\subsection{IBFE solution}

The diagram to be evaluated is:

\begin{equation}
G=%
\begin{minipage}{3.8cm} \includegraphics[scale=.7]
{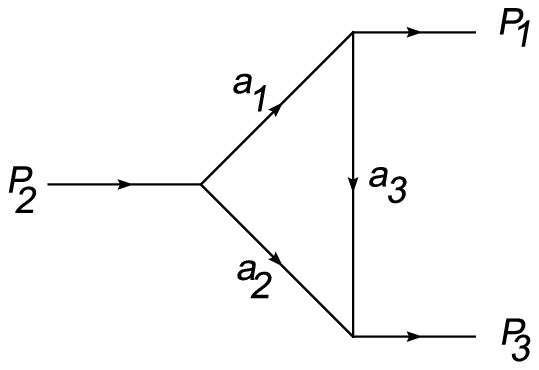} \end{minipage},
\end{equation}%
such that the propagator associated to the index one has a mass $m$, whereas
$P_{1}^{2}=P_{3}^{2}=0$ and $P_{2}^{2}=s$ \cite{EBo}. For this diagram then,
the equivalent configuration of invariants is described as follows:

\begin{equation}
G=\begin{minipage}{3.6cm} \includegraphics[scale=.7] {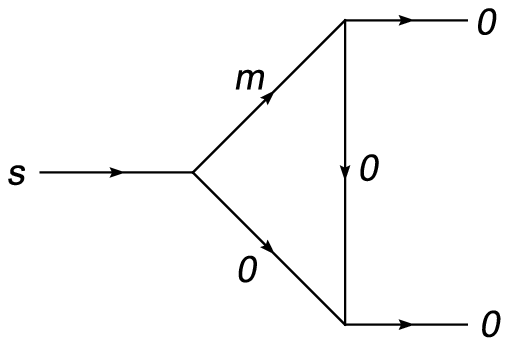}
\end{minipage}.  \label{ibfe16}
\end{equation}%
We will start by writing the corresponding Schwinger parametrization for
this topology which is given by

\begin{equation}
G=\dfrac{(-1)^{-D/2}}{\prod\nolimits_{j=1}^{3}\Gamma (a_{j})}%
\int\limits_{0}^{\infty }d\overrightarrow{x}\;\exp \left( x_{1}m^{2}\right)
\frac{\exp \left( -\dfrac{x_{1}x_{2}}{x_{1}+x_{2}+x_{3}}s\right) }{\left(
x_{1}+x_{2}+x_{3}\right) ^{D/2}},  \label{ibfe22}
\end{equation}%
where $d\overrightarrow{x}%
=x_{1}^{a_{1}-1}x_{2}^{a_{2}-1}x_{3}^{a_{3}-1}dx_{1}dx_{2}dx_{3}$. Some
algebra allows us to obtain the MRE \cite{IGoIBFE}\ form of $\left( \ref%
{ibfe22}\right) $:

\begin{equation}
G=\dfrac{(-1)^{-D/2}}{\prod\nolimits_{j=1}^{3}\Gamma (a_{j})}%
\sum\limits_{n_{1},...,n_{5}}\phi _{n_{1},...,n_{5}}\ \left( -m^{2}\right)
^{n_{1}}\left( s\right) ^{n_{2}}\dfrac{\prod\nolimits_{j=1}^{4}\Delta _{j}}{%
\Gamma (\frac{D}{2}+n_{2})},  \label{ibfe23}
\end{equation}%
where the factor $\phi _{n_{1},...,n_{5}}$, defined in \cite{IGoIBFE}%
\thinspace\ is given as follows:

\begin{equation}
\phi _{n_{1},...,n_{5}}=(-1)^{_{n_{1}+...+n_{5}}}\dfrac{1}{\Gamma
(n_{1}+1)...\Gamma (n_{5}+1)}.
\end{equation}%
The constraints are given by the identities

\begin{equation}
\left\{
\begin{array}{l}
\Delta _{1}=\left\langle \frac{D}{2}+n_{2}+n_{3}+n_{4}+n_{5}\right\rangle ,
\\
\Delta _{2}=\left\langle a_{1}+n_{1}+n_{2}+n_{3}\right\rangle , \\
\Delta _{3}=\left\langle a_{2}+n_{2}+n_{4}\right\rangle , \\
\Delta _{4}=\left\langle a_{3}+n_{5}\right\rangle .%
\end{array}%
\right.
\end{equation}%
Each constraint, according to the IBFE technique \cite{IGoIBFE}\, is
associated to the replacement of the integral, after expanding the
integrand, with the following symbol:

\begin{equation}
\int dx\;x^{\alpha _{1}+...+\alpha _{2}-1}\equiv \left\langle \alpha
_{1}+...+\alpha _{2}\right\rangle
\end{equation}%
which also emerged as a result from multinomial expansions that obey the
following rule \cite{IGoIBFE}:

\begin{equation}
\left( A_{1}+...+A_{\sigma }\right) ^{\pm \nu
}=\sum\limits_{n_{1}}...\sum\limits_{n_{\sigma }}\phi _{n_{1},...,n_{\sigma
}}\ A_{1}^{n_{1}}...A_{\sigma }^{n_{\sigma }}\frac{\left\langle \mp \nu
+n_{1}+...+n_{\sigma }\right\rangle }{\Gamma (\mp \nu )},
\end{equation}%
where the terms $A_{i}$ $\left( i=1,...,\sigma \right) $ and the exponent $%
\nu $ are quantities that can take arbitrary values.

\subsubsection{Solution for arbitrary indices}

The properties of the general solutions for Feynman diagrams imply the
existence of two interesting kinematical regions:

\paragraph{Solution in the region $\left\vert \dfrac{s}{m^{2}}\right\vert <1$%
}

\begin{equation}
G\left( \tfrac{s}{m^{2}}\right) =\eta \times \;_{2}F_{1}\left( \left.
\begin{array}{c}
\begin{array}{ccc}
a_{1}+a_{2}+a_{3}-\tfrac{D}{2} & , & a_{2}%
\end{array}
\\
\begin{array}{c}
\tfrac{D}{2}%
\end{array}%
\end{array}%
\right\vert \dfrac{s}{m^{2}}\right) ,  \label{ibfe13}
\end{equation}%
where the factor $\eta $ is defined as:

\begin{equation}
\eta =(-1)^{-\tfrac{D}{2}}\left( -m^{2}\right) ^{\tfrac{D}{2}%
-a_{1}-a_{2}-a_{3}}\dfrac{\Gamma \left( a_{1}+a_{2}+a_{3}-\tfrac{D}{2}%
\right) \Gamma \left( \tfrac{D}{2}-a_{2}-a_{3}\right) }{\Gamma \left(
a_{1}\right) \Gamma \left( \tfrac{D}{2}\right) }.
\end{equation}

\paragraph{Solution in the region $\left\vert \dfrac{m^{2}}{s}\right\vert <1$%
}

\begin{equation}
\begin{array}{ll}
G\left( \tfrac{m^{2}}{s}\right) = & \eta \times \;_{2}F_{1}\left( \left.
\begin{array}{c}
\begin{array}{ccc}
1+a_{1}+a_{2}+a_{3}-D & , & a_{1}+a_{2}+a_{3}-\tfrac{D}{2}%
\end{array}
\\
\begin{array}{c}
1+a_{1}+a_{3}-\tfrac{D}{2}%
\end{array}%
\end{array}%
\right\vert \dfrac{m^{2}}{s}\right) \\
&  \\
& +\;\underline{\eta }\times \;_{2}F_{1}\left( \left.
\begin{array}{c}
\begin{array}{ccc}
1+a_{2}-\tfrac{D}{2} & , & a_{2}%
\end{array}
\\
\begin{array}{c}
1+\tfrac{D}{2}-a_{1}-a_{3}%
\end{array}%
\end{array}%
\right\vert \dfrac{m^{2}}{s}\right) ,%
\end{array}
\label{ibfe14}
\end{equation}%
where the different factors are:

\begin{equation}
\eta =(-1)^{-\tfrac{D}{2}}\left( s\right) ^{\tfrac{D}{2}-a_{1}-a_{2}-a_{3}}%
\dfrac{\Gamma \left( a_{1}+a_{2}+a_{3}-\tfrac{D}{2}\right) \Gamma \left(
\tfrac{D}{2}-a_{1}-a_{3}\right) \Gamma \left( \tfrac{D}{2}%
-a_{2}-a_{3}\right) }{\Gamma \left( a_{1}\right) \Gamma \left( a_{2}\right)
\Gamma \left( D-a_{1}-a_{2}-a_{3}\right) },
\end{equation}

\begin{equation}
\underline{\eta }=(-1)^{-\tfrac{D}{2}}\left( s\right) ^{-a_{2}}\left(
-m^{2}\right) ^{\tfrac{D}{2}-a_{1}-a_{3}}\dfrac{\Gamma \left( a_{1}+a_{3}-%
\tfrac{D}{2}\right) \Gamma \left( \tfrac{D}{2}-a_{2}-a_{3}\right) }{\Gamma
\left( a_{1}\right) \Gamma \left( \tfrac{D}{2}-a_{2}\right) }.
\end{equation}

\subsubsection{Solution for unitary indices}

This situation is especially relevant, since when an arbitrary Feynman
diagram is computed, the indices associated to the propagators are normally
unitary, this case also being useful for numerical computations. Then, with $%
D=4-2\epsilon $, the Eqs. $\left( \ref{ibfe13}\right) $ and $\left( \ref%
{ibfe14}\right) $ are:

\paragraph{Solution in the region $\left\vert \dfrac{s}{m^{2}}\right\vert <1$%
}

\begin{equation}
G\left( \tfrac{s}{m^{2}}\right) =-(-1)^{\epsilon }\left( -m^{2}\right)
^{-1-\epsilon }\dfrac{\Gamma \left( 1+\epsilon \right) }{\epsilon \left(
1-\epsilon \right) }\;_{2}F_{1}\left( \left.
\begin{array}{c}
\begin{array}{ccc}
1+\epsilon & , & 1%
\end{array}
\\
\begin{array}{c}
2-\epsilon%
\end{array}%
\end{array}%
\right\vert \dfrac{s}{m^{2}}\right) .  \label{ibfe15}
\end{equation}

\paragraph{Solution in the region $\left\vert \dfrac{m^{2}}{s}\right\vert <1$%
}

\begin{equation}
G\left( \tfrac{m^{2}}{s}\right) =\eta \times \;_{1}F_{0}\left( \left.
\begin{array}{c}
\begin{array}{c}
2\epsilon%
\end{array}
\\
\begin{array}{c}
-%
\end{array}%
\end{array}%
\right\vert \dfrac{m^{2}}{s}\right) +\;\underline{\eta }\times
\;_{2}F_{1}\left( \left.
\begin{array}{c}
\begin{array}{ccc}
\epsilon & , & 1%
\end{array}
\\
\begin{array}{c}
1-\epsilon%
\end{array}%
\end{array}%
\right\vert \dfrac{m^{2}}{s}\right)
\end{equation}%
and the factors

\begin{equation}
\eta =(-1)^{\epsilon }\left( s\right) ^{-1-\epsilon }\dfrac{\Gamma \left(
1+\epsilon \right) \Gamma \left( 1-\epsilon \right) ^{2}}{\epsilon
^{2}\Gamma \left( 1-2\epsilon \right) },
\end{equation}

\begin{equation}
\underline{\eta }=-(-1)^{\epsilon }\left( s\right) ^{-1}\left( -m^{2}\right)
^{-\epsilon }\dfrac{\Gamma \left( 1+\epsilon \right) }{\epsilon ^{2}}.
\end{equation}

\subsection{Solution obtained by the application of IBFE+IBP}

The first step required for combining these two techniques is to eliminate
the mass from the diagram in $\left( \ref{ibfe16}\right) $. Once we have
done this, we can apply IBP to the massless triangle $\left( \ref{ibfe1}%
\right) $. For this purpose, we will use IBFE on the propagator that
contains this mass and expand it according to this technique, as it will be
shown next:

\begin{equation}
\dfrac{1}{\left( q^{2}-m^{2}\right) ^{\nu }}=\sum\limits_{n_{1},n_{2}}\phi
_{n_{1},n_{2}}\;\left( -m^{2}\right) ^{n_{1}}\left( q^{2}\right) ^{n_{2}}%
\dfrac{\left\langle \nu +n_{1}+n_{2}\right\rangle }{\Gamma (\nu )},
\end{equation}%
or in a graphically equivalent way :

\begin{equation}
\dfrac{1}{\left( q^{2}-m^{2}\right) ^{\nu }}=\sum\limits_{n_{1},n_{2}}\phi
_{n_{1},n_{2}}\;\left( -m^{2}\right) ^{n_{1}}\dfrac{\left\langle \nu
+n_{1}+n_{2}\right\rangle }{\Gamma (\nu )}\times
\begin{array}{c}
\left( -n_{2}\right) \\
\begin{minipage}{2.0cm} \includegraphics[scale=.7] {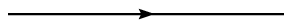}
\end{minipage}\;q \\
\end{array}%
.  \label{ibfe25}
\end{equation}%
Therefore, $\left( \ref{ibfe16}\right) $\ will be given by

\begin{equation}
\begin{minipage}{3.1cm} \includegraphics[scale=.7] {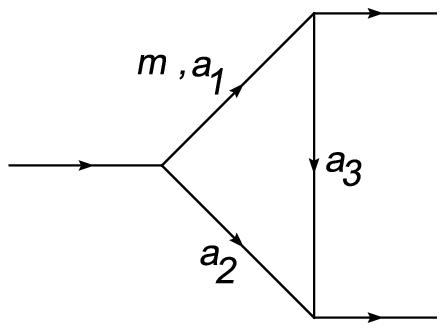}
\end{minipage}=\sum\limits_{n_{1},n_{2}}\phi _{n_{1},n_{2}}\ \left(
-m^{2}\right) ^{n_{1}}\dfrac{\left\langle a_{1}+n_{1}+n_{2}\right\rangle }{%
\Gamma (a_{1})}\times
\begin{minipage}{3.1cm} \includegraphics[scale=.7]
{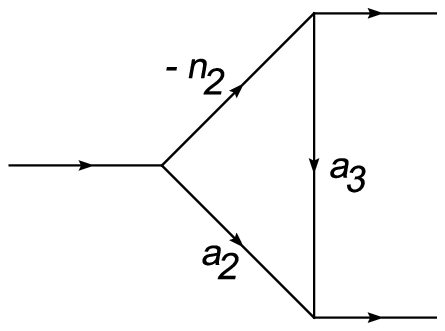} \end{minipage}.  \label{ibfe17}
\end{equation}%
The next step is obtained from the application of IBP to the triangle, once
the mass has been eliminated from the graph. However, before this, and
considering the invariants in $\left( \ref{ibfe16}\right) $\ we can rewrite
the triangle rule $\left( \ref{ibfe1}\right) $\ as

\begin{equation}
\begin{minipage}{3.1cm} \includegraphics[scale=.7] {rtriangle.eps}
\end{minipage}=\dfrac{1}{\left( D-a_{1}-a_{2}-2a_{3}\right) }\left[ \left(
a_{1}\right)
\begin{minipage}{3.1cm} \includegraphics[scale=.7]
{triangle1.eps} \end{minipage}+\left( a_{2}\right) \begin{minipage}{3.1cm}
\includegraphics[scale=.7] {triangle3.eps} \end{minipage}\right] .
\end{equation}%
In order to apply this rule, we will consider special values for the
indices: we take $a_{3}=1$ and keep arbitrary values for $a_{1}$ and $a_{2} $
($a_{4}=a_{5}=0$, since they are external lines). Then, the previous
expression can be written as:

\begin{equation}
\begin{minipage}{3.1cm} \includegraphics[scale=.7] {rtriangle.eps}
\end{minipage}=\dfrac{1}{\left( D-a_{1}-a_{2}-2\right) }\left[ \left(
a_{1}\right)
\begin{minipage}{3.1cm} \includegraphics[scale=.7]
{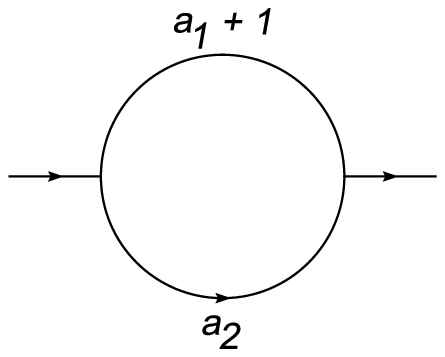} \end{minipage}+\left( a_{2}\right) \begin{minipage}{3.1cm}
\includegraphics[scale=.7] {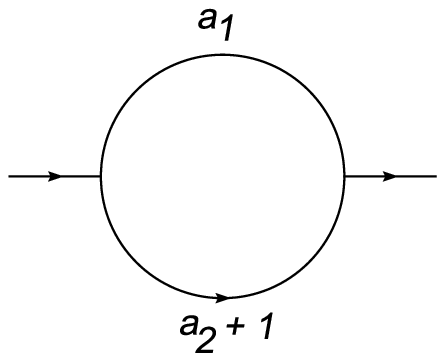} \end{minipage}\right] .
\label{ibfe18}
\end{equation}%
Applying $\left( \ref{ibfe18}\right) $ in $\left( \ref{ibfe17}\right) $ we
have:

\begin{equation}
\begin{array}{ll}
\begin{minipage}{3.1cm} \includegraphics[scale=.7] {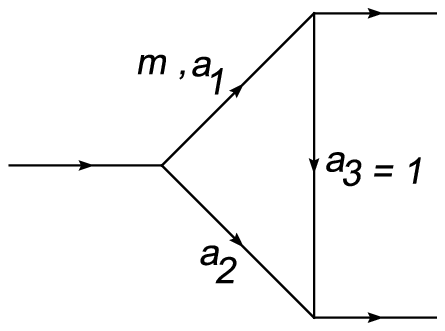}
\end{minipage} & =\sum\limits_{n_{1},n_{2}}\phi _{n_{1},n_{2}}\ \left(
-m^{2}\right) ^{n_{1}}\dfrac{\left\langle a_{1}+n_{1}+n_{2}\right\rangle }{%
\Gamma (a_{1})}\times
\begin{minipage}{3.1cm} \includegraphics[scale=.7]
{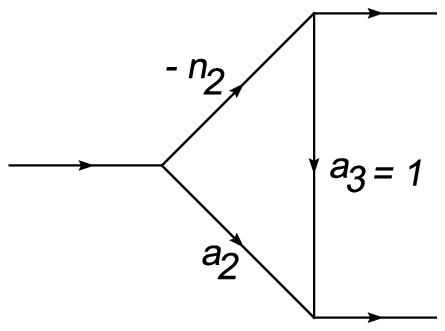} \end{minipage} \\
&  \\
& =\sum\limits_{n_{1},n_{2}}\phi _{n_{1},n_{2}}\ \left( -m^{2}\right)
^{n_{1}}\dfrac{\left\langle a_{1}+n_{1}+n_{2}\right\rangle }{\Gamma (a_{1})}%
\dfrac{1}{\left( D+n_{2}-a_{2}-2\right) } \\
&  \\
& \times \;\left[ \left( -n_{2}\right) \begin{minipage}{3.1cm}
\includegraphics[scale=.7] {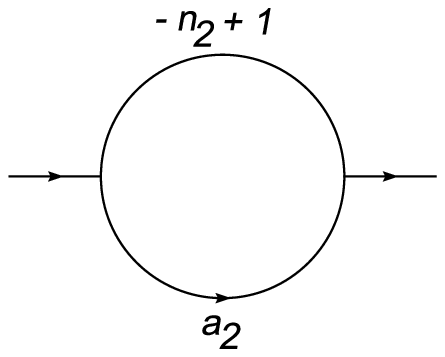} \end{minipage}+\left(
a_{2}\right)
\begin{minipage}{3.1cm} \includegraphics[scale=.7]
{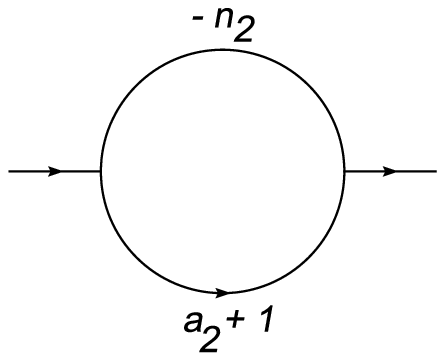} \end{minipage}\right] .%
\end{array}%
\end{equation}%
For simplicity, let us define:

\begin{equation}
A=-\sum\limits_{n_{1},n_{2}}\phi _{n_{1},n_{2}}\ \left( -m^{2}\right)
^{n_{1}}\dfrac{\left\langle a_{1}+n_{1}+n_{2}\right\rangle }{\Gamma (a_{1})}%
\dfrac{\Gamma (D+n_{2}-a_{2}-2)}{\Gamma (D+n_{2}-a_{2}-1)}\dfrac{\Gamma
(n_{2}+1)}{\Gamma (n_{2})}\times
\begin{minipage}{3.1cm} \includegraphics[scale=.7]
{tr_bub1_1.eps} \end{minipage},
\end{equation}%
and also:

\begin{equation}
B=\dfrac{\Gamma (a_{2}+1)}{\Gamma (a_{2})}\sum\limits_{n_{1},n_{2}}\phi
_{n_{1},n_{2}}\ \left( -m^{2}\right) ^{n_{1}}\dfrac{\left\langle
a_{1}+n_{1}+n_{2}\right\rangle }{\Gamma (a_{1})}\dfrac{\Gamma
(D+n_{2}-a_{2}-2)}{\Gamma (D+n_{2}-a_{2}-1)}\times \begin{minipage}{3.1cm}
\includegraphics[scale=.7] {tr_bub1_2.eps} \end{minipage},
\end{equation}%
where the identity $\sigma =\dfrac{\Gamma (\sigma +1)}{\Gamma (\sigma )}$
has been used to express everything in terms of Gamma functions.

To find the MRE's associated to the terms $A$ and $B$, we only need to
express the bubble diagrams in these equations through the following
identity \cite{IGo}:

\begin{equation}
\begin{array}{c}
\alpha \\
\begin{minipage}{3.1cm} \includegraphics[scale=.7] {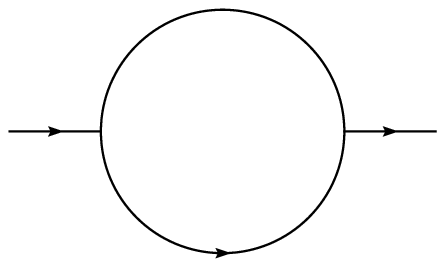}
\end{minipage}\;p \\
\beta%
\end{array}%
=\sum\limits_{n_{j}}\mathbf{G}_{A}(\alpha ,\beta ;\;n_{j})\times
\begin{array}{c}
\left( -n_{j}\right) \\
\begin{minipage}{2.0cm} \includegraphics[scale=.7] {propagador.eps}
\end{minipage}\;p \\
\end{array}%
,
\end{equation}%
where the one bubble function is given by

\begin{equation}
\mathbf{G}_{A}\left( \alpha ,\beta ;\;n_{j}\right) =\dfrac{(-1)^{-\frac{D}{2}%
}}{\Gamma (\alpha )\Gamma (\beta )}\sum\limits_{n_{(j+1)},n_{(j+2)}}\phi
_{n_{j},n_{(j+1)},n_{(j+2)}}\;\dfrac{\Delta _{1}\Delta _{2}\Delta _{3}}{%
\Gamma (\frac{D}{2}+n_{j})},
\end{equation}%
the constraints being $\left\{ \Delta _{i}\right\} $

\begin{equation}
\left\{
\begin{array}{l}
\Delta _{1}=\left\langle \frac{D}{2}+n_{j}+n_{(j+1)}+n_{(j+2)}\right\rangle ,
\\
\Delta _{2}=\left\langle \alpha +n_{j}+n_{(j+1)}\right\rangle , \\
\Delta _{3}=\left\langle \beta +n_{j}+n_{(j+2)}\right\rangle .%
\end{array}%
\right.
\end{equation}%
Through a careful replacement in $A$ and $B$ we find for $A$

\begin{equation}
\begin{array}{ll}
A= & -\dfrac{(-1)^{-\frac{D}{2}}}{\Gamma (a_{1})\Gamma (a_{2})}%
\sum\limits_{n_{1},..,n_{5}}\phi _{n_{1},..,n_{5}}\ \left( -m^{2}\right)
^{n_{1}}\left( s\right) ^{n_{3}}\dfrac{\Gamma (D+n_{2}-a_{2}-2)\Gamma
(n_{2}+1)}{\Gamma (D+n_{2}-a_{2}-1)\Gamma (n_{2})\Gamma (1-n_{2})\Gamma (%
\frac{D}{2}+n_{3})} \\
&  \\
& \times \;\left\langle a_{1}+n_{1}+n_{2}\right\rangle \left\langle \frac{D}{%
2}+n_{3}+n_{4}+n_{5}\right\rangle \left\langle
1-n_{2}+n_{3}+n_{4}\right\rangle \left\langle a_{2}+n_{3}+n_{5}\right\rangle%
\end{array}
\label{ibfe24}
\end{equation}%
and $B$ is given by

\begin{equation}
\begin{array}{ll}
B= & \dfrac{(-1)^{-\frac{D}{2}}}{\Gamma (a_{1})\Gamma (a_{2})}%
\sum\limits_{n_{1},..,n_{5}}\phi _{n_{1},..,n_{5}}\ \left( -m^{2}\right)
^{n_{1}}\left( s\right) ^{n_{3}}\dfrac{\Gamma (D+n_{2}-a_{2}-2)}{\Gamma
(D+n_{2}-a_{2}-1)\Gamma (-n_{2})\Gamma (\frac{D}{2}+n_{3})} \\
&  \\
& \times \;\left\langle a_{1}+n_{1}+n_{2}\right\rangle \left\langle \frac{D}{%
2}+n_{3}+n_{4}+n_{5}\right\rangle \left\langle
-n_{2}+n_{3}+n_{4}\right\rangle \left\langle
1+a_{2}+n_{3}+n_{5}\right\rangle .%
\end{array}%
\end{equation}%
Once we have obtained the MRE's of the terms $A$ and $B$, we proceed to find
the corresponding solutions according to the quotient $\left\vert \dfrac{%
m^{2}}{s}\right\vert $.

\subsubsection{Solutions from the term A}

\paragraph{Solution for the region $\left\vert \dfrac{m^{2}}{s}\right\vert
<1:$}

From the MRE $\left( \ref{ibfe24}\right) $ we get the following solution for
this kinematical region:

\begin{equation}
A_{1}\left( \tfrac{m^{2}}{s}\right) =\eta \times \;_{3}F_{2}\left( \left.
\begin{array}{c}
\begin{array}{ccccc}
2+a_{1}+a_{2}-D & , & 2+a_{1}+a_{2}-D & , & 1+a_{1}+a_{2}-\tfrac{D}{2}%
\end{array}
\\
\begin{array}{ccc}
3+a_{1}+a_{2}-D & , & 2+a_{1}-\tfrac{D}{2}%
\end{array}%
\end{array}%
\right\vert \dfrac{m^{2}}{s}\right) ,
\end{equation}%
where the factor $\eta $ is given by :

\begin{equation}
\eta =(-1)^{-\tfrac{D}{2}}\left( s\right) ^{\tfrac{D}{2}-a_{1}-a_{2}-1}%
\dfrac{\Gamma \left( a_{1}+a_{2}+1-\tfrac{D}{2}\right) \Gamma \left( \tfrac{D%
}{2}-a_{2}\right) \Gamma \left( \tfrac{D}{2}-a_{1}-1\right) \Gamma \left(
D-a_{1}-a_{2}-2\right) }{\Gamma \left( a_{1}\right) \Gamma \left(
a_{2}\right) \Gamma \left( D-a_{1}-a_{2}-1\right) ^{2}}
\end{equation}%
and also the term

\begin{equation}
A_{2}\left( \tfrac{m^{2}}{s}\right) =\eta \times \;_{3}F_{2}\left( \left.
\begin{array}{c}
\begin{array}{ccccc}
a_{2} & , & 1+a_{2}-\tfrac{D}{2} & , & 1+a_{2}-\tfrac{D}{2}%
\end{array}
\\
\begin{array}{ccc}
2+a_{2}-\tfrac{D}{2} & , & \tfrac{D}{2}-a_{1}%
\end{array}%
\end{array}%
\right\vert \dfrac{m^{2}}{s}\right) ,
\end{equation}%
where we have defined $\eta $ as

\begin{equation}
\eta =-(-1)^{-\tfrac{D}{2}}\left( s\right) ^{-a_{2}}\left( -m^{2}\right) ^{%
\tfrac{D}{2}-a_{1}-1}\dfrac{\Gamma \left( a_{1}+1-\tfrac{D}{2}\right) \Gamma
\left( \tfrac{D}{2}-1\right) \Gamma \left( \tfrac{D}{2}-2\right) \Gamma
\left( \tfrac{D}{2}-a_{2}-1\right) }{\Gamma \left( a_{1}\right) \Gamma
\left( \tfrac{D}{2}-a_{2}\right) \Gamma \left( 1-\tfrac{D}{2}\right) \Gamma
\left( \tfrac{D}{2}\right) }.
\end{equation}

\paragraph{Solution for the region $\left\vert \dfrac{m^{2}}{s}\right\vert
>1:$}

In this region we only have one term :

\begin{equation}
A_{3}\left( \tfrac{m^{2}}{s}\right) =\eta \times \;_{3}F_{2}\left( \left.
\begin{array}{c}
\begin{array}{ccccc}
a_{2} & , & 1+a_{1}+a_{2}-\tfrac{D}{2} & , & \tfrac{D}{2}-1%
\end{array}
\\
\begin{array}{ccc}
\tfrac{D}{2} & , & \tfrac{D}{2}%
\end{array}%
\end{array}%
\right\vert \dfrac{s}{m^{2}}\right) ,
\end{equation}%
where

\begin{equation}
\eta =-(-1)^{-\tfrac{D}{2}}\left( -m^{2}\right) ^{\tfrac{D}{2}-a_{1}-a_{2}-1}%
\dfrac{\Gamma \left( 1+a_{1}+a_{2}-\tfrac{D}{2}\right) \Gamma \left( \tfrac{D%
}{2}-a_{2}-1\right) \Gamma \left( \tfrac{D}{2}-1\right) \Gamma \left(
2+a_{2}-\tfrac{D}{2}\right) }{\Gamma \left( a_{1}\right) \Gamma \left(
1+a_{2}-\tfrac{D}{2}\right) \Gamma \left( \tfrac{D}{2}\right) ^{2}}.
\end{equation}

\subsubsection{Solutions from the term B}

\paragraph{Solution for the region $\left\vert \dfrac{m^{2}}{s}\right\vert
<1:$}

For $B$ we have as solution in this kinematical region

\begin{equation}
B_{1}\left( \tfrac{m^{2}}{s}\right) =\eta \times \;_{3}F_{2}\left( \left.
\begin{array}{c}
\begin{array}{ccccc}
2+a_{1}+a_{2}-D & , & 2+a_{1}+a_{2}-D & , & 1+a_{1}+a_{2}-\tfrac{D}{2}%
\end{array}
\\
\begin{array}{ccc}
3+a_{1}+a_{2}-D & , & 1+a_{1}-\tfrac{D}{2}%
\end{array}%
\end{array}%
\right\vert \dfrac{m^{2}}{s}\right) ,
\end{equation}%
being

\begin{equation}
\eta =(-1)^{-\tfrac{D}{2}}\left( s\right) ^{\tfrac{D}{2}-a_{1}-a_{2}-1}%
\dfrac{\Gamma \left( a_{1}+a_{2}+1-\tfrac{D}{2}\right) \Gamma \left( \tfrac{D%
}{2}-a_{2}-1\right) \Gamma \left( \tfrac{D}{2}-a_{1}\right) \Gamma \left(
D-a_{1}-a_{2}-2\right) }{\Gamma \left( a_{1}\right) \Gamma \left(
a_{2}\right) \Gamma \left( D-a_{1}-a_{2}-1\right) ^{2}}
\end{equation}%
and also the term

\begin{equation}
B_{2}\left( \tfrac{m^{2}}{s}\right) =\eta \times \;_{3}F_{2}\left( \left.
\begin{array}{c}
\begin{array}{ccccc}
1+a_{2} & , & 2+a_{2}-\tfrac{D}{2} & , & 2+a_{2}-\tfrac{D}{2}%
\end{array}
\\
\begin{array}{ccc}
3+a_{2}-\tfrac{D}{2} & , & \tfrac{D}{2}+1-a_{1}%
\end{array}%
\end{array}%
\right\vert \dfrac{m^{2}}{s}\right) ,
\end{equation}%
where

\begin{equation}
\eta =(-1)^{-\tfrac{D}{2}}\left( s\right) ^{-1-a_{2}}\left( -m^{2}\right) ^{%
\tfrac{D}{2}-a_{1}}\dfrac{\Gamma \left( a_{1}-\tfrac{D}{2}\right) \Gamma
\left( \tfrac{D}{2}-2-a_{2}\right) \Gamma \left( a_{2}+1\right) }{\Gamma
\left( a_{1}\right) \Gamma \left( a_{2}\right) \Gamma \left( \tfrac{D}{2}%
-1-a_{2}\right) }.
\end{equation}

\paragraph{Solution for the region $\left\vert \dfrac{m^{2}}{s}\right\vert
>1:$}

In this case we only get one hypergeometric function:

\begin{equation}
B_{3}\left( \tfrac{s}{m^{2}}\right) =\eta \times \;_{3}F_{2}\left( \left.
\begin{array}{c}
\begin{array}{ccccc}
1+a_{2} & , & 1+a_{1}+a_{2}-\tfrac{D}{2} & , & \tfrac{D}{2}-1%
\end{array}
\\
\begin{array}{ccc}
\tfrac{D}{2} & , & \tfrac{D}{2}%
\end{array}%
\end{array}%
\right\vert \dfrac{s}{m^{2}}\right) ,
\end{equation}%
where

\begin{equation}
\eta =(-1)^{-\tfrac{D}{2}}\left( -m^{2}\right) ^{\tfrac{D}{2}-a_{1}-a_{2}-1}%
\dfrac{\Gamma \left( 1+a_{1}+a_{2}-\tfrac{D}{2}\right) \Gamma \left( \tfrac{D%
}{2}-a_{2}-1\right) \Gamma \left( \tfrac{D}{2}-1\right) \Gamma \left(
1+a_{2}\right) }{\Gamma \left( a_{1}\right) \Gamma \left( a_{2}\right)
\Gamma \left( \tfrac{D}{2}\right) ^{2}}.
\end{equation}

\subsection{Comparison of the solutions for the region $\left\vert \dfrac{s}{%
m^{2}}\right\vert <1$ (unitary indices)}

In order to compare the solutions we have obtained with both techniques, we
will take the result for the kinematical region $\left\vert \dfrac{s}{m^{2}}%
\right\vert <1$.

For the case of unitary indices and $D=4-2\epsilon $, the combined technique
IBP+IBFE produces

\begin{equation}
G\left( \tfrac{s}{m^{2}}\right) =B_{3}\left( \tfrac{s}{m^{2}}\right)
+A_{3}\left( \tfrac{s}{m^{2}}\right) ,
\end{equation}%
i.e.

\begin{equation}
\begin{array}{ll}
G_{IBP+IBFE}\left( \tfrac{s}{m^{2}}\right) = & -\left( -1\right) ^{\epsilon
}\left( -m^{2}\right) ^{-1-\epsilon }\dfrac{\Gamma \left( 1+\epsilon \right)
}{\epsilon \left( 1-\epsilon \right) ^{2}}\left[ \;_{3}F_{2}\left( \left.
\begin{array}{c}
\begin{array}{ccccc}
2 & , & 1+\epsilon & , & 1-\epsilon%
\end{array}
\\
\begin{array}{ccc}
2-\epsilon & , & 2-\epsilon%
\end{array}%
\end{array}%
\right\vert \dfrac{s}{m^{2}}\right) \right. \\
&  \\
& -\epsilon \left. \;_{3}F_{2}\left( \left.
\begin{array}{c}
\begin{array}{ccccc}
1 & , & 1+\epsilon & , & 1-\epsilon%
\end{array}
\\
\begin{array}{ccc}
2-\epsilon & , & 2-\epsilon%
\end{array}%
\end{array}%
\right\vert \dfrac{s}{m^{2}}\right) \right] .%
\end{array}
\label{ibfe19}
\end{equation}%
In the same region of interest, the solution given in $\left( \ref{ibfe15}%
\right) $\thinspace\ obtained by a unique application of IBFE is notoriously
different, both from the amount of terms as well as from the structure. Let
us discuss this point more in detail:

\begin{equation}
G_{IBFE}\left( \tfrac{s}{m^{2}}\right) =-(-1)^{\epsilon }\left(
-m^{2}\right) ^{-1-\epsilon }\dfrac{\Gamma \left( 1+\epsilon \right) }{%
\epsilon \left( 1-\epsilon \right) }\;_{2}F_{1}\left( \left.
\begin{array}{c}
\begin{array}{ccc}
1+\epsilon & , & 1%
\end{array}
\\
\begin{array}{c}
2-\epsilon%
\end{array}%
\end{array}%
\right\vert \dfrac{s}{m^{2}}\right) .  \label{ibfe20}
\end{equation}%
We can do now comparisons to show the equivalence between both techniques
(see the Appendix for a general proof). A direct possibility is to analyze
the equivalence in certain limit cases, i.e. to compare, for $m=0$ the
solutions we obtained from the expressions valid in the kinematical region $%
\left\vert \dfrac{m^{2}}{s}\right\vert <1$ and to compare also the solutions
for $s=0$, this time considering the corresponding expressions for the
region $\left\vert \dfrac{m^{2}}{s}\right\vert >1$.

\subsubsection{Analytic equivalence in the limits $m\rightarrow 0$ and $%
s\rightarrow 0$}

\paragraph{\protect\underline{Solution for $m=0$}:}

In this limit, the Eq. $\left( \ref{ibfe19}\right) $ has the form

\begin{equation}
\left. G_{IBP+IBFE}\right\vert _{m=0}=A_{1}\left( 0\right) +B_{1}\left(
0\right) =2(-1)^{\epsilon }\left( s\right) ^{-1-\epsilon }\left[ \dfrac{%
\Gamma \left( 1+\epsilon \right) \Gamma \left( 1-\epsilon \right) ^{2}}{%
2\epsilon ^{2}\Gamma \left( 1-2\epsilon \right) }\right] ,
\end{equation}%
whereas $\left( \ref{ibfe20}\right) $

\begin{equation}
\left. G_{IBFE}\right\vert _{m=0}=(-1)^{\epsilon }\left( s\right)
^{-1-\epsilon }\dfrac{\Gamma \left( 1+\epsilon \right) \Gamma \left(
1-\epsilon \right) ^{2}}{\epsilon ^{2}\Gamma \left( 1-2\epsilon \right) }.
\end{equation}%
Some algebra allows us to show that effectively,

\begin{equation}
\left. G_{IBP+IBFE}\right\vert _{m=0}-\left. G_{IBFE}\right\vert _{m=0}=0.
\end{equation}

\paragraph{\protect\underline{Solution for $s=0$}:}

The Eq. $\left( \ref{ibfe19}\right) $ becomes now

\begin{equation}
\left. G_{IBP+IBFE}\right\vert _{s=0}=A_{3}\left( 0\right) +B_{3}\left(
0\right) =-\left( -1\right) ^{\epsilon }\left( -m^{2}\right) ^{-1-\epsilon }%
\dfrac{\Gamma \left( 1+\epsilon \right) }{\epsilon \left( 1-\epsilon \right)
}.
\end{equation}%
In the same way as $\left( \ref{ibfe20}\right) $, in this limit we have:

\begin{equation}
\left. G_{IBFE}\right\vert _{s=0}=-(-1)^{\epsilon }\left( -m^{2}\right)
^{-1-\epsilon }\dfrac{\Gamma \left( 1+\epsilon \right) }{\epsilon \left(
1-\epsilon \right) }.
\end{equation}%
In this case both solutions are equivalent:

\begin{equation}
\left. G_{IBP+IBFE}\right\vert _{s=0}-\left. G_{IBFE}\right\vert _{s=0}=0.
\end{equation}%
According to the results we have obtained previously, we can say that
IBP+IBFE is a technique that can be used in an equivalent way to IBFE to
evaluate Feynman diagrams. Nevertheless, the former has some technical
advantages that extend the usefulness of IBFE beyond the family of diagrams,
where it can ideally be applied and which is described in \cite{IGoIBFE}.

\section{A two loop application}

\qquad As an example in a more complex scenario, we will use now the
combined method IBP+IBFE for computing the correction to the propagator
given by two loops and five propagators. It is possible to evaluate such a
diagram for the massless case and with integer indices in their propagators
iterating the use of IBP (if it is necessary), transforming the original
topology into a sum of topologies which, for the unitary indices case,
generates more simple graphs which can be reduced loop by loop in terms of
bubble diagrams. In this work, we will evaluate this diagram taking now one
massive propagator, mass $M$ \cite{Broad}, with index $a_{1}$, as we can see
in the following graphic equation:

\begin{equation}
G=\begin{minipage}{6.1cm} \includegraphics[scale=.7] {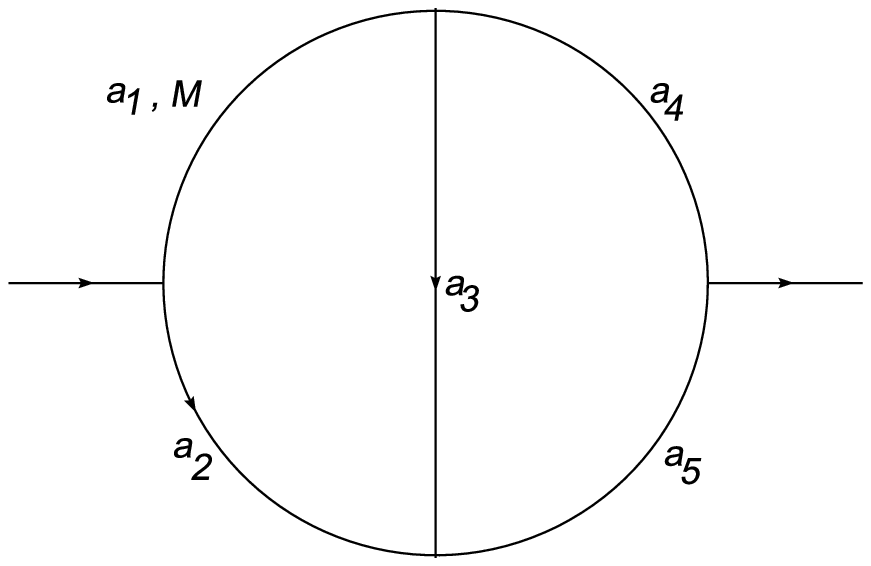} \end{minipage}%
\;p.  \label{ibfe26}
\end{equation}%
The solutions to this diagram are functions of the variables $%
\Longrightarrow $ $\left( \dfrac{p^{2}}{M^{2}}\right) $ or $\left( \dfrac{%
M^{2}}{p^{2}}\right) $ according to the kinematical region of interest. We
impose the following conditions to evaluate the diagram:

\begin{itemize}
\item The indices $a_{3},a_{4},$ and $a_{5}$ are integer quantities. In
particular, we will have $a_{3}=a_{4}=a_{5}=1$. This assumption does not
imply a loss of generality, since the indices in the loop integrals are
unitary.

\item We will consider arbitrary indices $a_{1}$ and $a_{2}$. This will
allow us not only the possibility of taking several values for the indices,
but also to implement topological variations of the diagram by inserting
corrections to the propagators, having then more loops, or by taking equal
or different masses for these propagators. The last point is related to the
fact that the extraction of the mass from a certain propagator generates a
massless propagator with a variable index, as it is shown in the Eq. $\left( %
\ref{ibfe25}\right) $.

\item Let us assume the masses $m_{2}=m_{3}=m_{4}=m_{5}=0$ and $m_{1}=M$.
\end{itemize}

\bigskip

Let us then consider the loop integral for this graph:

\begin{equation}
G=\int \frac{d^{D}q_{1}}{i\pi ^{D/2}}\frac{d^{D}q_{2}}{i\pi ^{D/2}}\frac{1}{%
(q_{1}^{2}-M^{2})^{a_{1}}}\frac{1}{\left[ \left( q_{1}+p\right) ^{2}\right]
^{a_{2}}}\frac{1}{\left( q_{2}+q_{1}\right) ^{2}}\frac{1}{q_{2}^{2}}\frac{1}{%
\left( p-q_{2}\right) ^{2}}.  \label{ibfe2}
\end{equation}%
The idea is to extract the mass $M$ from the integral in such a way that the
resulting integral corresponds to a massless diagram and then, in a second
step, to apply the triangle rule $\left( \ref{ibfe1}\right) $.

This is the place where IBFE can be used. This technique allows us to find
the MRE of the propagator that contains the mass , i.e.%
\begin{equation}
\frac{1}{(q_{1}^{2}-M^{2})^{a_{1}}}=\sum\limits_{n_{1},n_{2}}\phi
_{n_{1},n_{2}}\left( q_{1}^{2}\right) ^{n_{1}}\left( -M^{2}\right) ^{n_{2}}%
\dfrac{\left\langle a_{1}+n_{1}+n_{2}\right\rangle }{\Gamma \left(
a_{1}\right) }.
\end{equation}%
By replacing then in the integral $\left( \ref{ibfe2}\right) $, we get the
following expansion (In this expansion, we must keep in mind that the sum
indices are not necessarily integers):%
\begin{equation}
G=\sum\limits_{n_{1},n_{2}}\phi _{n_{1},n_{2}}\;\left( -M^{2}\right) ^{n_{2}}%
\dfrac{\left\langle a_{1}+n_{1}+n_{2}\right\rangle }{\Gamma \left(
a_{1}\right) }\int \dfrac{d^{D}q_{1}}{i\pi ^{D/2}}\dfrac{d^{D}q_{2}}{i\pi
^{D/2}}\dfrac{1}{(q_{1}^{2})^{-n_{1}}}\dfrac{1}{\left[ \left( q_{1}+p\right)
^{2}\right] ^{a_{2}}}\dfrac{1}{\left( q_{2}+q_{1}\right) ^{2}}\dfrac{1}{%
q_{2}^{2}}\dfrac{1}{\left( p-q_{2}\right) ^{2}}.
\end{equation}%
We can represent also in a graphic way this result in the following way:

\begin{equation}
G=\sum\limits_{n_{1},n_{2}}\phi _{n_{1},n_{2}}\;\left( -M^{2}\right) ^{n_{2}}%
\dfrac{\left\langle a_{1}+n_{1}+n_{2}\right\rangle }{\Gamma \left(
a_{1}\right) }\times
\begin{minipage}{6.1cm} \includegraphics[scale=.7]
{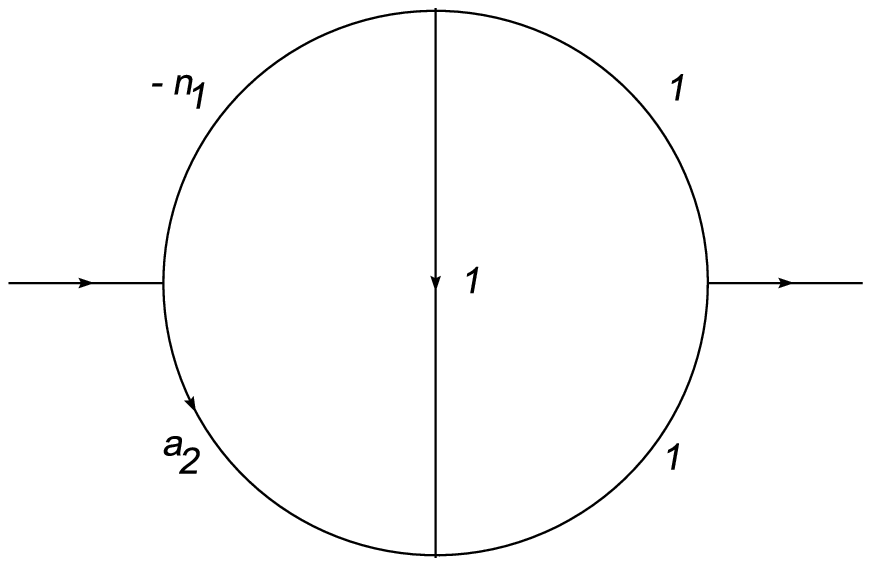} \end{minipage}\;p.  \label{ibfe8}
\end{equation}%
At this point, we can apply the triangle rule to the resulting topology. We
have to mention that the multiregion expansion of the massive propagator
contains in its structure the expansion for the cases of small and big
masses simultaneously, which is precisely what we need to compute this
diagram in all of the interesting kinematical zones.

\subsection{Application of the triangle rule}

To apply the formula $\left( \ref{ibfe1}\right) $, we will consider the left
triangle of the diagram in $\left( \ref{ibfe8}\right) $. As a result of
this, we get in a simple way the following expression for $G$:

\begin{equation}
\begin{array}{ll}
G= & \sum\limits_{n_{1},n_{2}}\phi _{n_{1},n_{2}}\;\left( -M^{2}\right)
^{n_{2}}\dfrac{\left\langle a_{1}+n_{1}+n_{2}\right\rangle }{\Gamma \left(
a_{1}\right) }\dfrac{1}{\left( D+n_{1}-a_{2}-2\right) } \\
&  \\
& \times \;\left[ \left( -n_{1}\right) \begin{minipage}{6.1cm}
\includegraphics[scale=.7] {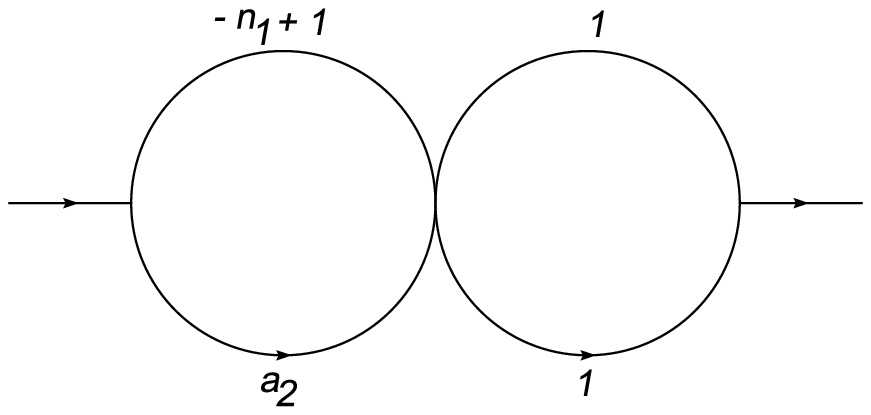} \end{minipage}-\left( -n_{1}\right) %
\begin{minipage}{5.9cm} \includegraphics[scale=.7] {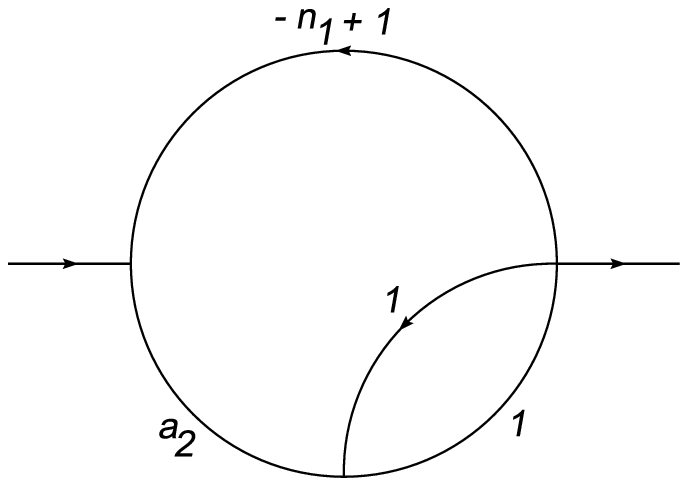} \end{minipage}%
\right. \\
&  \\
& \left. +\left( a_{2}\right) \begin{minipage}{6.1cm}
\includegraphics[scale=.7] {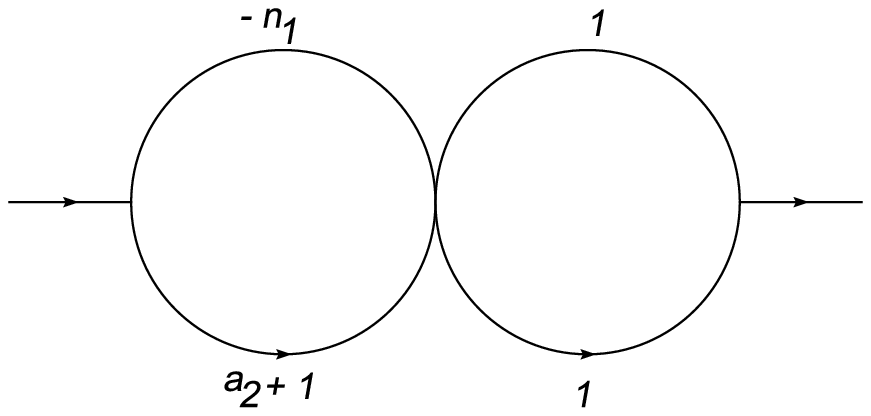} \end{minipage}-\left( a_{2}\right) %
\begin{minipage}{4.8cm} \includegraphics[scale=.7] {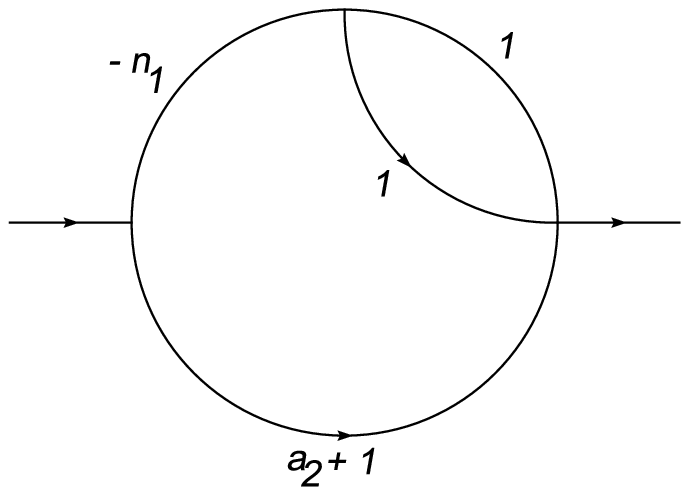} \end{minipage}%
\right] .%
\end{array}
\label{ibfe3}
\end{equation}%
We will express all the coefficients (only those which contain some sum
index) of the previous equation in terms of Gamma functions. This will be
useful to form the resulting hypergeometric functions. We use the following
identity:

\begin{equation}
\Gamma \left( \alpha +1\right) =\alpha \Gamma \left( \alpha \right)
\Longrightarrow \alpha =\frac{\Gamma \left( \alpha +1\right) }{\Gamma \left(
\alpha \right) },
\end{equation}%
with this, we may replace in $\left( \ref{ibfe3}\right) $\ the sum index
dependent coefficients

\begin{equation}
\left( -n_{1}\right) =\frac{\Gamma \left( -n_{1}+1\right) }{\Gamma \left(
-n_{1}\right) },
\end{equation}%
and for the denominator,

\begin{equation}
\frac{1}{\left( D+n_{1}-a_{2}-2\right) }=\frac{\Gamma \left(
D+n_{1}-a_{2}-2\right) }{\Gamma \left( D+n_{1}-a_{2}-1\right) }.
\end{equation}

\subsection{Analytical Solution}

According to the Eq. $\left( \ref{ibfe3}\right) $, we can write the original
diagram as a sum of four terms, i.e.

\begin{equation}
G=A_{1}+A_{2}+A_{3}+A_{4},
\end{equation}%
which are defined as

\begin{equation}
\begin{array}{ll}
A_{1}= & \sum\limits_{n_{1},n_{2}}\phi _{n_{1},n_{2}}\;\left( -M^{2}\right)
^{n_{2}}\dfrac{\left\langle a_{1}+n_{1}+n_{2}\right\rangle }{\Gamma \left(
a_{1}\right) }\dfrac{\Gamma \left( D+n_{1}-a_{2}-2\right) }{\Gamma \left(
D+n_{1}-a_{2}-1\right) }\dfrac{\Gamma \left( -n_{1}+1\right) }{\Gamma \left(
-n_{1}\right) } \\
&  \\
& \times \;%
\begin{minipage}{6.1cm} \includegraphics[scale=.7]
{g1.eps} \end{minipage}\;p,%
\end{array}%
\end{equation}

\begin{equation}
\begin{array}{ll}
A_{2}= & \sum\limits_{n_{1},n_{2}}\phi _{n_{1},n_{2}}\;\left( -M^{2}\right)
^{n_{2}}\dfrac{\left\langle a_{1}+n_{1}+n_{2}\right\rangle }{\Gamma \left(
a_{1}\right) }\dfrac{\Gamma \left( D+n_{1}-a_{2}-2\right) }{\Gamma \left(
D+n_{1}-a_{2}-1\right) }\dfrac{\Gamma \left( n_{1}+1\right) }{\Gamma \left(
n_{1}\right) } \\
&  \\
& \times \;%
\begin{minipage}{4.8cm} \includegraphics[scale=.7]
{g2.eps} \end{minipage}\;p,%
\end{array}%
\end{equation}

\begin{equation}
\begin{array}{ll}
A_{3}= & \left( a_{2}\right) \sum\limits_{n_{1},n_{2}}\phi
_{n_{1},n_{2}}\;\left( -M^{2}\right) ^{n_{2}}\dfrac{\left\langle
a_{1}+n_{1}+n_{2}\right\rangle }{\Gamma \left( a_{1}\right) }\dfrac{\Gamma
\left( D+n_{1}-a_{2}-2\right) }{\Gamma \left( D+n_{1}-a_{2}-1\right) } \\
&  \\
& \times \;%
\begin{minipage}{6.1cm} \includegraphics[scale=.7]
{g3.eps} \end{minipage}\;p,%
\end{array}%
\end{equation}%
and, finally,

\begin{equation}
\begin{array}{ll}
A_{4}= & \left( -a_{2}\right) \sum\limits_{n_{1},n_{2}}\phi
_{n_{1},n_{2}}\;\left( -M^{2}\right) ^{n_{2}}\dfrac{\left\langle
a_{1}+n_{1}+n_{2}\right\rangle }{\Gamma \left( a_{1}\right) }\dfrac{\Gamma
\left( D+n_{1}-a_{2}-2\right) }{\Gamma \left( D+n_{1}-a_{2}-1\right) } \\
&  \\
& \times \;%
\begin{minipage}{4.8cm} \includegraphics[scale=.7]
{g4.eps} \end{minipage}\;p.%
\end{array}%
\end{equation}%
The resulting topologies in each term $A_{i}\left( i=1,2,3,4\right) $ are
appropriate to compute them directly with the IBFE integration technique.
With this technique, we will be able to get the analytical solution in the
possible kinematical regimes where each term $A_{i}$ contributes. In what
follows, it is necessary to get the equivalent complete MRE for each one of
these terms to obtain then the analytical solution. We have two ways to get
this expansion: the first one is to expand the parametrical integral that
corresponds to the complete diagram present in each term $A_{i}$, and the
second one, which is the most straightforward procedure, is to use the idea
of modular reduction of the diagram \cite{IGo}, that allows us to find very
rapidly the MRE of each one of the diagrams as a product of MRE of their
constituent individual loops. These are known as loop functions. We will use
this last form to start with the analysis of the remnant diagrams in each
term $A_{i}$. We will present the complete calculation procedure for the
first term of $\left( \ref{ibfe3}\right) $, $A_{1}$, and then the other
terms will be presented in a summarized way.

\subsubsection{Complete solution of $A_{1}$}

To find the solution from this term, we need first the MRE associated to the
diagram. We will use the idea of loop function representations \cite{IGo} to
obtain the expansion which has the form :

\begin{equation}
\begin{minipage}{6.1cm} \includegraphics[scale=.7] {g1.eps} \end{minipage}%
\;p=\sum\limits_{n_{3}}G_{A}(-n_{1}+1,a_{2};\;n_{3})\sum%
\limits_{n_{6}}G_{A}(1,1;\;n_{6})\times
\begin{array}{c}
\left( -n_{3}-n_{6}\right) \\
\begin{minipage}{2.0cm} \includegraphics[scale=.7] {propagador.eps}
\end{minipage}\;p \\
\end{array}%
,  \label{ibfe4}
\end{equation}%
where the loop function $G_{A}$ corresponds to the MRE of each loop, or part
of the diagram, and which has the following MRE:

\begin{equation}
G_{A}\left( a_{j},a_{k};\;n_{i}\right) =\dfrac{(-1)^{-\frac{D}{2}}}{\Gamma
(a_{j})\Gamma (a_{k})}\sum\limits_{n_{(i+1)},n_{(i+2)}}\phi
_{n_{i},n_{(i+1)},n_{(i+2)}}\;\dfrac{\Delta _{1}\Delta _{2}\Delta _{3}}{%
\Gamma (\frac{D}{2}+n_{i})}.  \label{ibfe11}
\end{equation}%
In this case the constraints $\left\{ \Delta _{i}\right\} $ are given by the
following expressions:%
\begin{equation}
\left\{
\begin{array}{l}
\Delta _{1}=\left\langle \frac{D}{2}+n_{i}+n_{(i+1)}+n_{(i+2)}\right\rangle ,
\\
\Delta _{2}=\left\langle a_{j}+n_{i}+n_{(i+1)}\right\rangle , \\
\Delta _{3}=\left\langle a_{k}+n_{i}+n_{(i+2)}\right\rangle .%
\end{array}%
\right.
\end{equation}%
Clearly, we have to replace the pictorial representation of the propagator
by its corresponding algebraic expression:

\begin{equation}
\begin{array}{c}
\alpha \\
\begin{minipage}{2.0cm} \includegraphics[scale=.7] {propagador.eps}
\end{minipage}\;p \\
\end{array}%
=\dfrac{1}{\left( p^{2}\right) ^{\alpha }}.
\end{equation}%
Some algebra allows us to find the MRE of the diagram in $\left( \ref{ibfe4}%
\right) $:

\begin{equation}
\begin{array}{ll}
\begin{minipage}{6.1cm} \includegraphics[scale=.7] {g1.eps} \end{minipage}%
\;p= & \dfrac{(-1)^{-D}}{\Gamma (-n_{1}+1)\Gamma \left( a_{2}\right) } \\
&  \\
& \times \sum\limits_{n_{3},...,n_{8}}\phi _{n_{3},...,n_{8}}\;\left(
p^{2}\right) ^{n_{3}+n_{6}}\dfrac{\Delta _{1}^{\left( 1\right) }...\Delta
_{6}^{\left( 1\right) }}{\Gamma (\frac{D}{2}+n_{3})\Gamma (\frac{D}{2}+n_{6})%
},%
\end{array}%
\end{equation}%
whose constraints are determined as follows:

\begin{equation}
\left\{
\begin{array}{l}
\Delta _{1}^{\left( 1\right) }=\left\langle \frac{D}{2}+n_{3}+n_{4}+n_{5}%
\right\rangle , \\
\Delta _{2}^{\left( 1\right) }=\left\langle
-n_{1}+1+n_{3}+n_{4}\right\rangle , \\
\Delta _{3}^{\left( 1\right) }=\left\langle a_{2}+n_{3}+n_{5}\right\rangle ,
\\
\Delta _{4}^{\left( 1\right) }=\left\langle \frac{D}{2}+n_{6}+n_{7}+n_{8}%
\right\rangle , \\
\Delta _{5}^{\left( 1\right) }=\left\langle 1+n_{6}+n_{7}\right\rangle , \\
\Delta _{6}^{\left( 1\right) }=\left\langle 1+n_{6}+n_{8}\right\rangle .%
\end{array}%
\right.
\end{equation}%
Finally, by introducing the necessarily replacements, we have obtained the
MRE for the first term of the solution of the diagram $G$. We have added a
superindex in the constraints. This only refers to the fact that these
constraints are associated to the first term of $G$, $A_{1}$. So, we get
then the MRE for this term:

\begin{equation}
A_{1}=\dfrac{(-1)^{-D}}{\Gamma \left( a_{1}\right) \Gamma \left(
a_{2}\right) }\sum\limits_{n_{1},...,n_{8}}\phi _{n_{1},...,n_{8}}\;\left(
-M^{2}\right) ^{n_{2}}\left( p^{2}\right) ^{n_{3}+n_{6}}\dfrac{\Gamma \left(
D+n_{1}-a_{2}-2\right) }{\Gamma \left( D+n_{1}-a_{2}-1\right) \Gamma \left(
-n_{1}\right) }\dfrac{\Delta _{1}^{\left( 1\right) }...\Delta _{6}^{\left(
1\right) }\Delta }{\Gamma (\tfrac{D}{2}+n_{3})\Gamma (\tfrac{D}{2}+n_{6})},
\label{ibfe5}
\end{equation}%
where we have also defined an extra constraint associated to the MRE of the
massive propagator:

\begin{equation}
\Delta =\left\langle a_{1}+n_{1}+n_{2}\right\rangle .
\end{equation}%
Starting from the expression $\left( \ref{ibfe5}\right) $, it is possible to
find the solutions in the two possible kinematical regions $\left\vert
\dfrac{p^{2}}{M^{2}}\right\vert <1$ and $\left\vert \dfrac{M^{2}}{p^{2}}%
\right\vert <1$, where the solutions for each kinematical region can be
divided into groups, as a sum of generalized hypergeometric functions and
where the sets will be defined by the terms $A_{1}\left( \tfrac{p^{2}}{M^{2}}%
\right) $ and $A_{1}\left( \tfrac{M^{2}}{p^{2}}\right) $, respectively.

We will show now the analytical solutions we found for the term $A_{1}$
according to the corresponding kinematical regions:

\paragraph{Solution in the region $\left\vert \dfrac{p^{2}}{M^{2}}%
\right\vert <1$}

\begin{equation}
A_{1}\left( \tfrac{p^{2}}{M^{2}}\right) =\eta \times \;_{3}F_{2}\left(
\left.
\begin{array}{c}
\begin{array}{ccccc}
1+a_{1}+a_{2}-\tfrac{D}{2} & , & a_{2} & , & \tfrac{D}{2}-1%
\end{array}
\\
\begin{array}{ccc}
\tfrac{D}{2} & , & \tfrac{D}{2}%
\end{array}%
\end{array}%
\right\vert \dfrac{p^{2}}{M^{2}}\right)
\end{equation}%
where the factor $\eta $\ is defined as:

\begin{equation}
\eta =(-1)^{-D}\left( p^{2}\right) ^{\tfrac{D}{2}-2}\left( -M^{2}\right) ^{%
\tfrac{D}{2}-1-a_{1}-a_{2}}\dfrac{\Gamma \left( 1+a_{1}+a_{2}-\tfrac{D}{2}%
\right) \Gamma \left( \tfrac{D}{2}-a_{2}\right) \Gamma \left( 2-\tfrac{D}{2}%
\right) \Gamma \left( \tfrac{D}{2}-1\right) ^{3}}{\Gamma \left( a_{1}\right)
\Gamma \left( D-2\right) \Gamma \left( \tfrac{D}{2}\right) ^{2}}.
\end{equation}

\paragraph{Solution in the region $\left\vert \dfrac{M^{2}}{p^{2}}%
\right\vert <1$}

\begin{equation}
\begin{array}{ll}
A_{1}\left( \tfrac{M^{2}}{p^{2}}\right) = & \eta \times \;_{3}F_{2}\left(
\left.
\begin{array}{c}
\begin{array}{ccccc}
2+a_{1}+a_{2}-D & , & 2+a_{1}+a_{2}-D & , & 1+a_{1}+a_{2}-\tfrac{D}{2}%
\end{array}
\\
\begin{array}{ccc}
2+a_{1}-\tfrac{D}{2} & , & 3+a_{1}+a_{2}-D%
\end{array}%
\end{array}%
\right\vert \dfrac{M^{2}}{p^{2}}\right) \\
&  \\
& +\;\underline{\eta }\times \;_{3}F_{2}\left( \left.
\begin{array}{c}
\begin{array}{ccccc}
1+a_{2}-\tfrac{D}{2} & , & 1+a_{2}-\tfrac{D}{2} & , & a_{2}%
\end{array}
\\
\begin{array}{ccc}
\tfrac{D}{2}-a_{1} & , & 2+a_{2}-\tfrac{D}{2}%
\end{array}%
\end{array}%
\right\vert \dfrac{M^{2}}{p^{2}}\right)%
\end{array}%
\end{equation}%
being the factors

\begin{equation}
\begin{array}{ll}
\eta = & (-1)^{-D}\left( p^{2}\right) ^{D-3-a_{1}-a_{2}}\dfrac{\Gamma \left(
1+a_{1}+a_{2}-\tfrac{D}{2}\right) \Gamma \left( \tfrac{D}{2}-a_{2}\right)
\Gamma \left( 2-\tfrac{D}{2}\right) \Gamma \left( \tfrac{D}{2}%
-a_{1}-1\right) }{\Gamma \left( a_{1}\right) \Gamma \left( a_{2}\right)
\Gamma \left( D-2\right) } \\
&  \\
& \times \;\dfrac{\Gamma \left( D-a_{1}-a_{2}-2\right) \Gamma \left( \tfrac{D%
}{2}-1\right) ^{2}}{\Gamma \left( D-a_{1}-a_{2}-1\right) ^{2}},%
\end{array}%
\end{equation}%
and

\begin{equation}
\underline{\eta }=(-1)^{-D}\left( p^{2}\right) ^{\tfrac{D}{2}-2-a_{2}}\left(
-M^{2}\right) ^{\tfrac{D}{2}-1-a_{1}}\dfrac{\Gamma \left( 1+a_{1}-\tfrac{D}{2%
}\right) \Gamma \left( \tfrac{D}{2}-a_{2}-1\right) \Gamma \left( 2-\tfrac{D}{%
2}\right) \Gamma \left( \tfrac{D}{2}-1\right) ^{2}}{\Gamma \left(
a_{1}\right) \Gamma \left( D-2\right) \Gamma \left( \tfrac{D}{2}%
-a_{2}\right) }.
\end{equation}%
Nevertheless, the solution of the diagram $G$ is by no means complete, since
we have to consider also the terms that can be generated from the MRE
associated to $\left\{ A_{2},A_{3},A_{4}\right\} $. Then, we have to add
them algebraically, according to the kinematical region of interest. i.e.:

\begin{equation}
G=\left\{
\begin{array}{lll}
A_{1}\left( \tfrac{p^{2}}{M^{2}}\right) +A_{2}\left( \tfrac{p^{2}}{M^{2}}%
\right) +A_{3}\left( \tfrac{p^{2}}{M^{2}}\right) +A_{4}\left( \tfrac{p^{2}}{%
M^{2}}\right) &  & ,\text{ for }M^{2}>\left\vert p^{2}\right\vert \\
&  &  \\
A_{1}\left( \tfrac{M^{2}}{p^{2}}\right) +A_{2}\left( \tfrac{M^{2}}{p^{2}}%
\right) +A_{3}\left( \tfrac{M^{2}}{p^{2}}\right) +A_{4}\left( \tfrac{M^{2}}{%
p^{2}}\right) &  & ,\text{ for }\left\vert p^{2}\right\vert >M^{2}.%
\end{array}%
\right.
\end{equation}%
We will proceed in an analogous way, as we did for the calculation of $A_{1}$%
, for the other terms $A_{i}$:

\subsubsection{Contributions to the solution generated by $A_{2}$}

For the diagram that appears in $A_{2}$, the MRE is obtained as l:

\begin{equation}
\begin{minipage}{4.8cm} \includegraphics[scale=.7] {g2.eps} \end{minipage}%
\;p=\sum\limits_{n_{3}}G_{A}(1,1;\;n_{3})\sum%
\limits_{n_{6}}G_{A}(-n_{1}+1,a_{2}-n_{3};\;n_{6})\times
\begin{array}{c}
-n_{6} \\
\begin{minipage}{2.0cm} \includegraphics[scale=.7] {propagador.eps}
\end{minipage}\;p \\
\end{array}%
,
\end{equation}%
and we use then the expression $\left( \ref{ibfe11}\right) $\ to replace the
one-loop functions $G_{A}$ in the expansion of this graph, getting in an
explicit way the MRE of the term $A_{2}$:

\begin{equation}
\begin{array}{ll}
A_{2}= & \dfrac{(-1)^{-D}}{\Gamma \left( a_{1}\right) }\sum%
\limits_{n_{1},...,n_{8}}\phi _{n_{1},...,n_{8}}\;\left( -M^{2}\right)
^{n_{2}}\left( p^{2}\right) ^{n_{6}}\dfrac{\Gamma \left(
D+n_{1}-a_{2}-2\right) \Gamma \left( n_{1}+1\right) }{\Gamma \left(
D+n_{1}-a_{2}-1\right) \Gamma \left( n_{1}\right) } \\
&  \\
& \times \dfrac{\Delta _{1}^{\left( 2\right) }...\Delta _{6}^{\left(
2\right) }\Delta }{\Gamma (\tfrac{D}{2}+n_{3})\Gamma (\tfrac{D}{2}%
+n_{6})\Gamma \left( -n_{1}+1\right) \Gamma (a_{2}-n_{3})},%
\end{array}%
\end{equation}%
where the constraints correspond to the following expressions:

\begin{equation}
\left\{
\begin{array}{l}
\Delta _{1}^{\left( 2\right) }=\left\langle \frac{D}{2}+n_{3}+n_{4}+n_{5}%
\right\rangle , \\
\Delta _{2}^{\left( 2\right) }=\left\langle 1+n_{3}+n_{4}\right\rangle , \\
\Delta _{3}^{\left( 2\right) }=\left\langle 1+n_{3}+n_{5}\right\rangle , \\
\Delta _{4}^{\left( 2\right) }=\left\langle \frac{D}{2}+n_{6}+n_{7}+n_{8}%
\right\rangle , \\
\Delta _{5}^{\left( 2\right) }=\left\langle
-n_{1}+1+n_{6}+n_{7}\right\rangle , \\
\Delta _{6}^{\left( 2\right) }=\left\langle
a_{2}-n_{3}+n_{6}+n_{8}\right\rangle .%
\end{array}%
\right.
\end{equation}%
With this information, we can find in this case the solutions for both
relevant kinematical regions:

\paragraph{Solution in the region $\left\vert \dfrac{p^{2}}{M^{2}}%
\right\vert <1$}

\begin{equation}
A_{2}\left( \tfrac{p^{2}}{M^{2}}\right) =\eta \times \;_{3}F_{2}\left(
\left.
\begin{array}{c}
\begin{array}{ccccc}
3+a_{1}+a_{2}-D & , & 2+a_{2}-\tfrac{D}{2} & , & 1%
\end{array}
\\
\begin{array}{ccc}
\tfrac{D}{2} & , & 2%
\end{array}%
\end{array}%
\right\vert \dfrac{p^{2}}{M^{2}}\right) ,
\end{equation}%
where the factor $\eta $\ is given by the following expression:

\begin{equation}
\eta =(-1)^{-D}\left( -M^{2}\right) ^{D-3-a_{1}-a_{2}}\dfrac{\Gamma \left(
D-3-a_{2}\right) \Gamma \left( 3+a_{1}+a_{2}-D\right) \Gamma \left(
4+a_{2}-D\right) \Gamma \left( 2-\tfrac{D}{2}\right) \Gamma \left( \tfrac{D}{%
2}-1\right) ^{2}}{\Gamma \left( a_{1}\right) \Gamma \left( D-2\right) \Gamma
\left( \tfrac{D}{2}\right) \Gamma \left( 3+a_{2}-D\right) }.
\end{equation}

\paragraph{Solution in the region $\left\vert \dfrac{M^{2}}{p^{2}}%
\right\vert <1$}

\begin{equation}
\begin{array}{ll}
A_{2}\left( \tfrac{M^{2}}{p^{2}}\right) = & \eta \times \;_{2}F_{1}\left(
\left.
\begin{array}{c}
\begin{array}{ccc}
4+a_{1}+a_{2}-\tfrac{3D}{2} & , & 2+a_{1}+a_{2}-D%
\end{array}
\\
\begin{array}{c}
2+a_{1}-\tfrac{D}{2}%
\end{array}%
\end{array}%
\right\vert \dfrac{M^{2}}{p^{2}}\right) \\
&  \\
& +\;\underline{\eta }\times \;_{2}F_{1}\left( \left.
\begin{array}{c}
\begin{array}{ccc}
3+a_{2}-D & , & 1+a_{2}-\tfrac{D}{2}%
\end{array}
\\
\begin{array}{c}
\tfrac{D}{2}-a_{1}%
\end{array}%
\end{array}%
\right\vert \dfrac{M^{2}}{p^{2}}\right) ,%
\end{array}%
\end{equation}%
where the factors are defined as

\begin{equation}
\begin{array}{ll}
\eta = & (-1)^{-D}\left( p^{2}\right) ^{D-3-a_{1}-a_{2}}\dfrac{\Gamma \left(
3+a_{1}+a_{2}-D\right) \Gamma \left( D-a_{2}-2\right) \Gamma \left( 2-\tfrac{%
D}{2}\right) \Gamma \left( \tfrac{D}{2}-a_{1}-1\right) }{\Gamma \left(
a_{1}+1\right) \Gamma \left( -a_{1}\right) \Gamma \left( D-2\right) } \\
&  \\
& \times \dfrac{\Gamma \left( D-a_{1}-a_{2}-2\right) \Gamma \left( \tfrac{D}{%
2}-1\right) ^{2}\Gamma \left( 1-a_{1}\right) }{\Gamma \left(
D-a_{1}-a_{2}-1\right) \Gamma \left( \tfrac{3D}{2}-a_{1}-a_{2}-3\right)
\Gamma \left( 2+a_{2}-\tfrac{D}{2}\right) },%
\end{array}%
\end{equation}

\begin{equation}
\underline{\eta }=(-1)^{-D}\left( p^{2}\right) ^{\tfrac{D}{2}-2-a_{2}}\left(
-M^{2}\right) ^{\tfrac{D}{2}-1-a_{1}}\dfrac{\Gamma \left( 1+a_{1}-\tfrac{D}{2%
}\right) \Gamma \left( \tfrac{D}{2}-a_{2}-1\right) \Gamma \left( 2-\tfrac{D}{%
2}\right) ^{2}\Gamma \left( \tfrac{D}{2}-1\right) ^{3}}{\Gamma \left(
a_{1}\right) \Gamma \left( D-2\right) \Gamma \left( \tfrac{D}{2}%
-a_{2}\right) \Gamma \left( \tfrac{D}{2}\right) \Gamma \left( 1-\tfrac{D}{2}%
\right) }.
\end{equation}

\subsubsection{Contributions to the solution generated by $A_{3}$}

We will follow the same procedure for getting this term, i.e. to find the
MRE of $A_{3}$ and their corresponding analytical solutions:

\begin{equation}
\begin{minipage}{6.1cm} \includegraphics[scale=.7] {g3.eps} \end{minipage}%
\;p=\sum\limits_{n_{3}}G_{A}(-n_{1},a_{2}+1;\;n_{3})\sum%
\limits_{n_{6}}G_{A}(1,1;\;n_{6})\times
\begin{array}{c}
\left( -n_{3}-n_{6}\right) \\
\begin{minipage}{2.0cm} \includegraphics[scale=.7] {propagador.eps}
\end{minipage}\;p \\
\end{array}%
,
\end{equation}

\begin{equation}
A_{3}=\dfrac{(-1)^{-D}}{\Gamma \left( a_{1}\right) \Gamma \left(
a_{2}\right) }\sum\limits_{n_{1},...,n_{8}}\phi _{n_{1},...,n_{8}}\;\left(
-M^{2}\right) ^{n_{2}}\left( p^{2}\right) ^{n_{3}+n_{6}}\dfrac{\Gamma \left(
D+n_{1}-a_{2}-2\right) }{\Gamma \left( D+n_{1}-a_{2}-1\right) }\dfrac{\Delta
_{1}^{\left( 3\right) }...\Delta _{6}^{\left( 3\right) }\Delta }{\Gamma (%
\tfrac{D}{2}+n_{3})\Gamma (\tfrac{D}{2}+n_{6})\Gamma \left( -n_{1}\right) },
\end{equation}%
with the constraints:

\begin{equation}
\left\{
\begin{array}{l}
\Delta _{1}^{\left( 3\right) }=\left\langle \frac{D}{2}+n_{3}+n_{4}+n_{5}%
\right\rangle , \\
\Delta _{2}^{\left( 3\right) }=\left\langle -n_{1}+n_{3}+n_{4}\right\rangle ,
\\
\Delta _{3}^{\left( 3\right) }=\left\langle a_{2}+1+n_{3}+n_{5}\right\rangle
, \\
\Delta _{4}^{\left( 3\right) }=\left\langle \frac{D}{2}+n_{6}+n_{7}+n_{8}%
\right\rangle , \\
\Delta _{5}^{\left( 3\right) }=\left\langle 1+n_{6}+n_{7}\right\rangle , \\
\Delta _{6}^{\left( 3\right) }=\left\langle 1+n_{6}+n_{8}\right\rangle .%
\end{array}%
\right.
\end{equation}

\paragraph{Solution in the region $\left\vert \dfrac{p^{2}}{M^{2}}%
\right\vert <1$}

\begin{equation}
A_{3}\left( \tfrac{p^{2}}{M^{2}}\right) =\eta \times \;_{3}F_{2}\left(
\left.
\begin{array}{c}
\begin{array}{ccccc}
1+a_{1}+a_{2}-\tfrac{D}{2} & , & \tfrac{D}{2}-1 & , & 1+a_{2}%
\end{array}
\\
\begin{array}{ccc}
\tfrac{D}{2} & , & \tfrac{D}{2}%
\end{array}%
\end{array}%
\right\vert \dfrac{p^{2}}{M^{2}}\right) ,
\end{equation}

\begin{equation}
\begin{array}{ll}
\eta = & (-1)^{-D}\left( -M^{2}\right) ^{\tfrac{D}{2}-1-a_{1}-a_{2}}\left(
p^{2}\right) ^{\tfrac{D}{2}-2} \\
&  \\
& \times \;\dfrac{\Gamma \left( 1+a_{1}+a_{2}-\tfrac{D}{2}\right) \Gamma
\left( \tfrac{D}{2}-a_{2}-1\right) \Gamma \left( 1+a_{2}\right) \Gamma
\left( 2-\tfrac{D}{2}\right) \Gamma \left( \tfrac{D}{2}-1\right) ^{3}}{%
\Gamma \left( a_{1}\right) \Gamma \left( a_{2}\right) \Gamma \left(
D-2\right) \Gamma \left( \tfrac{D}{2}\right) ^{2}.}.%
\end{array}%
\end{equation}

\paragraph{Solution in the region $\left\vert \dfrac{M^{2}}{p^{2}}%
\right\vert <1$}

\begin{equation}
\begin{array}{ll}
A_{3}\left( \tfrac{M^{2}}{p^{2}}\right) = & \eta \times \;_{3}F_{2}\left(
\left.
\begin{array}{c}
\begin{array}{ccccc}
2+a_{1}+a_{2}-D & , & 2+a_{1}+a_{2}-D & , & 1+a_{1}+a_{2}-\tfrac{D}{2}%
\end{array}
\\
\begin{array}{ccc}
1+a_{1}-\tfrac{D}{2} & , & 3+a_{1}+a_{2}-D%
\end{array}%
\end{array}%
\right\vert \dfrac{M^{2}}{p^{2}}\right) \\
&  \\
& +\;\underline{\eta }\times \;_{3}F_{2}\left( \left.
\begin{array}{c}
\begin{array}{ccccc}
2+a_{2}-\tfrac{D}{2} & , & 2+a_{2}-\tfrac{D}{2} & , & 1+a_{2}%
\end{array}
\\
\begin{array}{ccc}
1-a_{1}+\tfrac{D}{2} & , & 3+a_{2}-\tfrac{D}{2}%
\end{array}%
\end{array}%
\right\vert \dfrac{M^{2}}{p^{2}}\right) ,%
\end{array}%
\end{equation}%
where the different factors are given by

\begin{equation}
\begin{array}{ll}
\eta = & (-1)^{-D}\left( p^{2}\right) ^{D-3-a_{1}-a_{2}}\dfrac{\Gamma \left(
D-a_{1}-a_{2}-2\right) \Gamma \left( \frac{D}{2}-a_{2}-1\right) \Gamma
\left( 2-\tfrac{D}{2}\right) \Gamma \left( \tfrac{D}{2}-1\right) ^{2}}{%
\Gamma \left( a_{1}\right) \Gamma \left( a_{2}\right) \Gamma \left(
D-2\right) } \\
&  \\
& \times \;\dfrac{\Gamma \left( \tfrac{D}{2}-a_{1}\right) \Gamma \left(
1+a_{1}+a_{2}-\frac{D}{2}\right) }{\Gamma \left( D-a_{1}-a_{2}-1\right) ^{2}}%
,%
\end{array}%
\end{equation}

\begin{equation}
\underline{\eta }=(-1)^{-D}\left( p^{2}\right) ^{\tfrac{D}{2}-3-a_{2}}\left(
-M^{2}\right) ^{\tfrac{D}{2}-a_{1}}\dfrac{\Gamma \left( a_{1}-\tfrac{D}{2}%
\right) \Gamma \left( 1+a_{2}\right) \Gamma \left( 2-\tfrac{D}{2}\right)
\Gamma \left( \tfrac{D}{2}-a_{2}-2\right) \Gamma \left( \tfrac{D}{2}%
-1\right) ^{2}}{\Gamma \left( a_{1}\right) \Gamma \left( a_{1}\right) \Gamma
\left( D-2\right) \Gamma \left( \tfrac{D}{2}-a_{2}-1\right) }.
\end{equation}

\subsubsection{Contributions to the solution generated by $A_{4}$}

\begin{equation}
\begin{minipage}{4.8cm} \includegraphics[scale=.7] {g4.eps} \end{minipage}%
\;p=\sum\limits_{n_{3}}G_{A}(1,1;\;n_{3})\sum%
\limits_{n_{6}}G_{A}(-n_{1}-n_{3},a_{2}+1;\;n_{6})\times
\begin{array}{c}
-n_{6} \\
\begin{minipage}{2.0cm} \includegraphics[scale=.7] {propagador.eps}
\end{minipage}\;p \\
\end{array}%
,
\end{equation}

\begin{equation}
A_{4}=-\dfrac{(-1)^{-D}}{\Gamma \left( a_{1}\right) \Gamma \left(
a_{2}\right) }\sum\limits_{n_{1},..,n_{8}}\phi _{n_{1},..,n_{8}}\;\left(
-M^{2}\right) ^{n_{2}}\left( p^{2}\right) ^{n_{6}}\dfrac{\Gamma \left(
D+n_{1}-a_{2}-2\right) }{\Gamma \left( D+n_{1}-a_{2}-1\right) }\dfrac{\Delta
_{1}^{\left( 4\right) }...\Delta _{6}^{\left( 4\right) }\Delta }{\Gamma (%
\tfrac{D}{2}+n_{3})\Gamma (\tfrac{D}{2}+n_{6})\Gamma \left(
-n_{1}-n_{3}\right) },
\end{equation}

\begin{equation}
\left\{
\begin{array}{l}
\Delta _{1}^{\left( 4\right) }=\left\langle \frac{D}{2}+n_{3}+n_{4}+n_{5}%
\right\rangle , \\
\Delta _{2}^{\left( 4\right) }=\left\langle 1+n_{3}+n_{4}\right\rangle , \\
\Delta _{3}^{\left( 4\right) }=\left\langle 1+n_{3}+n_{5}\right\rangle , \\
\Delta _{4}^{\left( 4\right) }=\left\langle \frac{D}{2}+n_{6}+n_{7}+n_{8}%
\right\rangle , \\
\Delta _{5}^{\left( 4\right) }=\left\langle
-n_{1}-n_{3}+n_{6}+n_{7}\right\rangle , \\
\Delta _{6}^{\left( 4\right) }=\left\langle a_{2}+1+n_{6}+n_{8}\right\rangle
.%
\end{array}%
\right.
\end{equation}

\paragraph{Solution in the region $\left\vert \dfrac{p^{2}}{M^{2}}%
\right\vert <1$}

\begin{equation}
\begin{array}{ll}
A_{4}\left( \tfrac{p^{2}}{M^{2}}\right) = & \eta \times \;_{3}F_{2}\left(
\left.
\begin{array}{c}
\begin{array}{ccccc}
D-2 & , & a_{1} & , & \tfrac{D}{2}-1%
\end{array}
\\
\begin{array}{ccc}
D-a_{2}-1 & , & \tfrac{3D}{2}-a_{2}-3%
\end{array}%
\end{array}%
\right\vert \dfrac{p^{2}}{M^{2}}\right) \\
&  \\
& +\;\underline{\eta }\times \;_{4}F_{3}\left( \left.
\begin{array}{c}
\begin{array}{ccccccc}
2+a_{2}-\tfrac{D}{2} & , & 3+a_{2}+a_{1}-D & , & a_{2}+1 & , & 1%
\end{array}
\\
\begin{array}{ccccc}
\tfrac{D}{2} & , & 2 & , & 4-D+a_{2}%
\end{array}%
\end{array}%
\right\vert \dfrac{p^{2}}{M^{2}}\right) ,%
\end{array}%
\end{equation}%
being

\begin{equation}
\eta =(-1)^{-D}\left( p^{2}\right) ^{D-3-a_{2}}\left( -M^{2}\right) ^{-a_{1}}%
\dfrac{\Gamma \left( 3+a_{2}-D\right) \Gamma \left( \frac{D}{2}%
-a_{2}-1\right) \Gamma \left( D-a_{2}-2\right) \Gamma \left( \tfrac{D}{2}%
-1\right) ^{2}}{\Gamma \left( a_{2}\right) \Gamma \left( \frac{3D}{2}%
-a_{2}-3\right) \Gamma \left( D-a_{2}-1\right) },
\end{equation}

\begin{equation}
\underline{\eta }=(-1)^{-D}\left( -M^{2}\right) ^{D-3-a_{1}-a_{2}}\dfrac{%
\Gamma \left( D-a_{2}-3\right) \Gamma \left( 3+a_{2}+a_{1}-D\right) \Gamma
\left( 2-\tfrac{D}{2}\right) \Gamma \left( \tfrac{D}{2}-1\right) ^{2}\Gamma
\left( 1+a_{2}\right) }{\Gamma \left( a_{1}\right) \Gamma \left(
a_{2}\right) \Gamma \left( D-2\right) \Gamma \left( \tfrac{D}{2}\right) }.
\end{equation}

\paragraph{Solution in the region $\left\vert \dfrac{M^{2}}{p^{2}}%
\right\vert <1$}

\begin{equation}
\begin{array}{ll}
A_{4}\left( \tfrac{M^{2}}{p^{2}}\right) = & \eta \times \;_{3}F_{2}\left(
\left.
\begin{array}{c}
\begin{array}{ccccc}
4+a_{1}+a_{2}-\frac{3D}{2} & , & 2+a_{1}+a_{2}-D & , & a_{1}%
\end{array}
\\
\begin{array}{ccc}
2+a_{1}-\tfrac{D}{2} & , & 3+a_{1}-D%
\end{array}%
\end{array}%
\right\vert \dfrac{M^{2}}{p^{2}}\right) \\
&  \\
& +\;\underline{\eta }\times \;_{3}F_{2}\left( \left.
\begin{array}{c}
\begin{array}{ccccc}
2+a_{2}-\tfrac{D}{2} & , & D-2 & , & a_{2}%
\end{array}
\\
\begin{array}{ccc}
D-a_{1}-1 & , & \tfrac{D}{2}%
\end{array}%
\end{array}%
\right\vert \dfrac{M^{2}}{p^{2}}\right) ,%
\end{array}%
\end{equation}%
with

\begin{equation}
\begin{array}{ll}
\eta = & (-1)^{-D}\left( p^{2}\right) ^{D-3-a_{1}-a_{2}}\dfrac{\Gamma \left(
\tfrac{D}{2}-a_{2}-1\right) \Gamma \left( D-a_{1}-2\right) \Gamma \left(
D-a_{1}-a_{2}-2\right) }{\Gamma \left( a_{2}\right) \Gamma \left( D-2\right)
} \\
&  \\
& \times \dfrac{\Gamma \left( 2-\tfrac{D}{2}\right) \Gamma \left(
3+a_{1}+a_{2}-D\right) \Gamma \left( \tfrac{D}{2}-1\right) ^{2}}{\Gamma
\left( D-a_{1}-a_{2}-1\right) \Gamma \left( \frac{3D}{2}-a_{1}-a_{2}-3%
\right) \Gamma \left( 2+a_{1}-\frac{D}{2}\right) },%
\end{array}%
\end{equation}

\begin{equation}
\underline{\eta }=(-1)^{-D}\left( p^{2}\right) ^{-1-a_{2}}\left(
-M^{2}\right) ^{D-a_{1}-2}\dfrac{\Gamma \left( 2+a_{1}-D\right) \Gamma
\left( 2-\tfrac{D}{2}\right) \Gamma \left( \tfrac{D}{2}-1\right) ^{2}\Gamma
\left( 1+a_{2}\right) \Gamma \left( -a_{2}\right) }{\Gamma \left(
a_{1}\right) \Gamma \left( a_{2}\right) \Gamma \left( \tfrac{D}{2}\right)
\Gamma \left( 1-a_{2}\right) },
\end{equation}

\subsection{A particular case: Solution for $M=0$ with indices $%
a_{1}=a_{2}=1 $}

\qquad In what follows, we will test the solution we got for the diagram $%
\left( \ref{ibfe26}\right) $. For this, we will compute the massless case,
which has a trivial solution when integration by parts is used. In this
case, we have to consider the solution obtained by IBP+IBFE in the region $%
\left\vert \dfrac{M^{2}}{p^{2}}\right\vert <1$ and to take it's limit $M=0$,
i.e.

\begin{equation}
G_{\left( M=0\right) }=\left[ A_{1}\left( \tfrac{M^{2}}{p^{2}}\right)
+A_{2}\left( \tfrac{M^{2}}{p^{2}}\right) +A_{3}\left( \tfrac{M^{2}}{p^{2}}%
\right) +A_{4}\left( \tfrac{M^{2}}{p^{2}}\right) \right] _{M=0}.
\end{equation}%
It is easy to verify that in this limit only four terms will survive. Using
some algebra, we can get the following result:

\begin{equation}
\begin{array}{ll}
G_{IBFE_{\left( M=0\right) }}= & 2\;(-1)^{-D}\left( p^{2}\right) ^{D-5}%
\dfrac{\Gamma \left( \tfrac{D}{2}-2\right) \Gamma \left( \tfrac{D}{2}%
-1\right) ^{2}\Gamma \left( 2-\tfrac{D}{2}\right) \Gamma \left( D-4\right) }{%
\Gamma \left( D-2\right) } \\
&  \\
& \times \;\left[ \dfrac{\Gamma \left( \tfrac{D}{2}-1\right) \Gamma \left( 3-%
\tfrac{D}{2}\right) }{\Gamma \left( D-3\right) ^{2}}-\dfrac{\Gamma \left(
5-D\right) }{\Gamma \left( \frac{3D}{2}-5\right) \Gamma \left( 3-\frac{D}{2}%
\right) }\right] .%
\end{array}
\label{ibfe6}
\end{equation}%
On the other side, the conditions under which we now have to evaluate the
diagram are such that the solution can be found in a conventional way, i.e.
applying IBP through the triangle rule. So, we find

\begin{equation}
\begin{minipage}{2.9cm} \includegraphics[scale=.7] {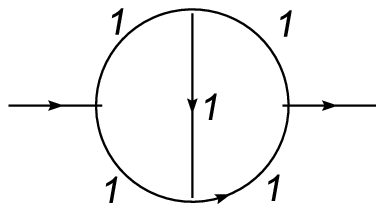}
\end{minipage}=\frac{2}{\left( D-4\right) }\left[ \begin{minipage}{3.1cm}
\includegraphics[scale=.7] {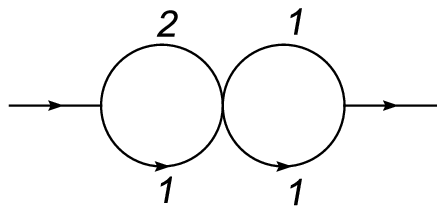} \end{minipage}\;-\;%
\begin{minipage}{3.1cm} \includegraphics[scale=.7] {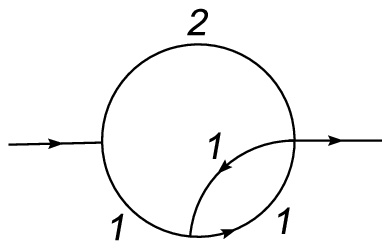}
\end{minipage}\right] ,
\end{equation}%
and we can rewrite this as

\begin{equation}
G_{IBP_{\left( M=0\right) }}=\frac{2}{\left( D-4\right) }\;G\left(
1,1\right) \left[ G\left( 2,1\right) -G\left( 2,3-\tfrac{D}{2}\right) \right]
\;\left( p^{2}\right) ^{D-5},
\end{equation}%
where the factors $G(\alpha ,\beta )$ are given by

\begin{equation}
G(\alpha ,\beta )=(-1)^{-\frac{D}{2}}\dfrac{\Gamma (\alpha +\beta -\frac{D}{2%
})\Gamma (\frac{D}{2}-\alpha )\Gamma (\frac{D}{2}-\beta )}{\Gamma (\alpha
)\Gamma (\beta )\Gamma (D-\alpha -\beta )}.
\end{equation}%
Finally, we obtain the following result for this two-loop diagram:

\begin{equation}
\begin{array}{ll}
G_{IBP_{\left( M=0\right) }}= & (-1)^{-D}\left( p^{2}\right) ^{D-5}\dfrac{2}{%
\left( D-4\right) }\dfrac{\Gamma \left( 2-\tfrac{D}{2}\right) \Gamma \left(
\tfrac{D}{2}-1\right) ^{2}\Gamma \left( \tfrac{D}{2}-2\right) }{\Gamma
\left( D-2\right) } \\
&  \\
& \times \left[ \dfrac{\Gamma \left( \tfrac{D}{2}-1\right) \Gamma \left( 3-%
\tfrac{D}{2}\right) }{\Gamma \left( D-3\right) }-\dfrac{\Gamma \left(
5-D\right) \Gamma \left( D-3\right) }{\Gamma \left( \frac{3D}{2}-5\right)
\Gamma \left( 3-\frac{D}{2}\right) }\right] .%
\end{array}
\label{ibfe7}
\end{equation}%
The solutions $\left( \ref{ibfe6}\right) $ and $\left( \ref{ibfe7}\right) $
seem to be apparently different, but by using the properties of the Gamma
function and some algebra, it is possible to show that both terms are equal,
i.e.

\begin{equation}
G_{IBP_{\left( M=0\right) }}-G_{IBFE_{\left( M=0\right) }}=0.
\end{equation}%
This result gives support to the most general solution for this diagram in
the case of one massive propagator.

\section{Conclusions and comments}

\qquad It is not always possible to combine the IBFE technique with another
conventional integration method. For example, let us evaluate the one-loop
diagram where one of the propagators has a mass $M$:

\begin{equation}
G=\int \frac{d^{D}k}{i\pi ^{D/2}}\frac{1}{(k^{2}-M^{2})^{a_{1}}\left[
(p-k)^{2}\right] ^{a_{2}}}.  \label{ibfe9}
\end{equation}%
As a first step, we will take out the mass $M$ of the massive propagator in
such a way that we may associate the resulting integral to a massless
diagram. We need to express this propagator in terms of a MRE, i.e.

\begin{equation}
\frac{1}{(k^{2}-M^{2})^{a_{1}}}=\sum\limits_{n_{1},n_{2}}\phi
_{n_{1},n_{2}}\left( k^{2}\right) ^{n_{1}}\left( -M^{2}\right) ^{n_{2}}%
\dfrac{\left\langle a_{1}+n_{1}+n_{2}\right\rangle }{\Gamma \left(
a_{1}\right) },
\end{equation}%
finding, when replacing this result in $\left( \ref{ibfe9}\right) $,

\begin{equation}
G=\sum\limits_{n_{1},n_{2}}\phi _{n_{1},n_{2}}\left( -M^{2}\right) ^{n_{2}}%
\dfrac{\left\langle a_{1}+n_{1}+n_{2}\right\rangle }{\Gamma \left(
a_{1}\right) }\int \frac{d^{D}k}{i\pi ^{D/2}}\frac{1}{(k^{2})^{-n_{1}}\left[
(p-k)^{2}\right] ^{a_{2}}}.  \label{ibfe12}
\end{equation}%
If the idea is to use only IBFE , the next step would be to find the
corresponding MRE for the resulting loop integral. However, instead of doing
this we will evaluate directly, and in a conventional way, the corresponding
integral. This integral is elementary and represents a massless bubble
diagram whose solution is given by

\begin{equation}
\int \frac{d^{D}k}{i\pi ^{D/2}}\frac{1}{(k^{2})^{-n_{1}}\left[ (p-k)^{2}%
\right] ^{a_{2}}}=(-1)^{-\frac{D}{2}}\dfrac{\Gamma (-n_{1}+a_{2}-\frac{D}{2}%
)\Gamma (\frac{D}{2}+n_{1})\Gamma (\frac{D}{2}-a_{2})}{\Gamma (-n_{1})\Gamma
(a_{2})\Gamma (D+n_{1}-a_{2})}\dfrac{1}{(p^{2})^{-n_{1}+a_{2}-\frac{D}{2}}}.
\end{equation}%
We replace this result in $\left( \ref{ibfe12}\right) $ getting the
following MRE for $G$:

\begin{equation}
G=(-1)^{-\frac{D}{2}}\dfrac{\Gamma (\frac{D}{2}-a_{2})}{\Gamma \left(
a_{1}\right) \Gamma (a_{2})}(p^{2})^{\frac{D}{2}-a_{2}}\sum%
\limits_{n_{1},n_{2}}\phi _{n_{1},n_{2}}\left( -M^{2}\right)
^{n_{2}}(p^{2})^{n_{1}}\left\langle a_{1}+n_{1}+n_{2}\right\rangle \dfrac{%
\Gamma (-n_{1}+a_{2}-\frac{D}{2})\Gamma (\frac{D}{2}+n_{1})}{\Gamma
(-n_{1})\Gamma (D+n_{1}-a_{2})}.  \label{ibfe10}
\end{equation}%
According to the prescription of IBFE, we notice immediately that the
maximum number of possible terms we can extract to form the solution is

$\left(
\begin{array}{c}
2 \\
1%
\end{array}%
\right) =2$. In fact, one of these terms corresponds to an hypergeometric
series which is solution for the kinematical region $\left\vert \dfrac{M^{2}%
}{p^{2}}\right\vert <1$ and the other one would give us the solution for the
region $\left\vert \dfrac{p^{2}}{M^{2}}\right\vert <1$. However, this
solution is not complete since some terms are missed \cite{EBo, CAn}. In
fact, it remains to add one term to the corresponding solution in the region
$\left\vert \dfrac{M^{2}}{p^{2}}\right\vert <1$. The terms which can be
extracted $\left( \ref{ibfe10}\right) $ are correct, but they represent only
a part of the solution, and this is the problem.

We will find now the MRE for this diagram using IBFE as the only integration
method. Then, we will compare with the previous result shown in $\left( \ref%
{ibfe10}\right) $. Schwinger's parametric expansion of $\left( \ref{ibfe9}%
\right) $ is%
\begin{equation}
G=\dfrac{(-1)^{-\frac{D}{2}}}{\Gamma (a_{1})\Gamma (a_{2})}%
\int\limits_{0}^{\infty }dx_{1}dx_{2}\;x_{1}^{a_{1}-1}x_{2}^{a_{2}-1}\frac{%
\exp \left( x_{1}M^{2}\right) \exp \left( -\dfrac{x_{1}x_{2}}{x_{1}+x_{2}}%
p^{2}\right) }{\left( x_{1}+x_{2}\right) ^{\frac{D}{2}}}.
\end{equation}%
After some algebra, the following MRE for the diagram $G$ is obtained:

\begin{equation}
G=\dfrac{(-1)^{-\frac{D}{2}}}{\Gamma (a_{1})\Gamma (a_{2})}%
\sum\limits_{n_{1},..,n_{4}}\phi _{n_{1,..,}n_{4}}\ \dfrac{\left(
p^{2}\right) ^{n_{1}}\left( -M^{2}\right) ^{n_{2}}}{\Gamma (\frac{D}{2}%
+n_{1})}\prod\nolimits_{j=1}^{3}\Delta _{j}  \label{ibfe21}
\end{equation}%
where

\begin{equation}
\begin{array}{l}
\Delta _{1}=\left\langle \frac{D}{2}+n_{1}+n_{3}+n_{4}\right\rangle , \\
\Delta _{2}=\left\langle a_{1}+n_{1}+n_{2}+n_{3}\right\rangle , \\
\Delta _{3}=\left\langle a_{2}+n_{1}+n_{4}\right\rangle .%
\end{array}%
\end{equation}%
We observe that this time, the possible number of series representations
that can be extracted from $\left( \ref{ibfe21}\right) $\ is given by the
combinatory $\left(
\begin{array}{c}
4 \\
3%
\end{array}%
\right) =4$. This shows that when conventional techniques, combined with
IBFE, are used, the MRE for a certain diagram is more simple than the MRE
obtained only by means of IBFE. This means that the conventional techniques
do not generate sums or Kroneker deltas and therefore the number of
different ways for summing, using the constraints of the MRE, becomes also
smaller.

In general, the solutions that are obtained from these MRE's do not
correspond necessarily to the correct solution. They can be incomplete. In
summary, once we have chosen to use IBFE as an integration technique, the
whole resulting process must be done with this technique. However, to
simplify the problem, it is possible to apply other techniques before IBFE,
but not viceversa.

The exception to this rule is IBP, where the integration in terms of the
topological cancellation of one of the graph lines do not produce the
disappearance of terms in the solution. The explanation for this relies in
the fact that the IBP expresses the original diagram as a sum of diagrams.
In the example given at the beginning of this work, we used IBP in the
intermediate step of getting the MRE. The result was a sum of MRE's in which
each one contributed to the solution, with terms associated to the
corresponding interesting kinematical regions. The conclusion is that IBP is
a technique compatible with IBFE.

In this work, we have emphasized the role of the triangle identity and the
fact that, under certain particular mass distributions, it is possible to
evaluate a series of diagrams with the procedure described above. For
example,

\begin{equation*}
\begin{minipage}{9.4cm} \includegraphics[scale=.7] {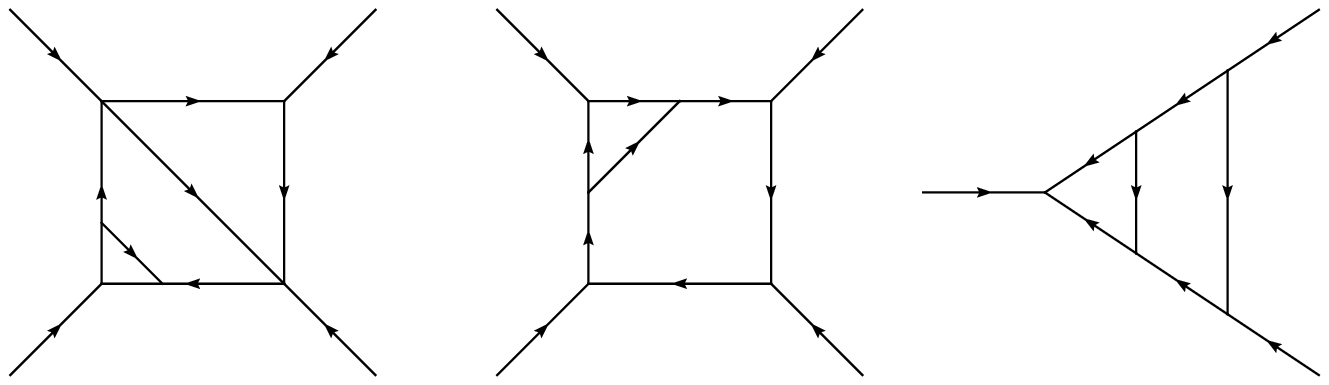} \end{minipage}
\end{equation*}%
Although these examples do not necessarily generate a hypergeometric series
of one variable, the procedure can be directly applied to these topologies.
Another option for generalizing is the insertion of bubbles in the
propagators, with arbitrary indices, which can be massless or massive. As an
example, let us consider the following case:

\begin{equation*}
\begin{minipage}{5.7cm} \includegraphics[scale=.6] {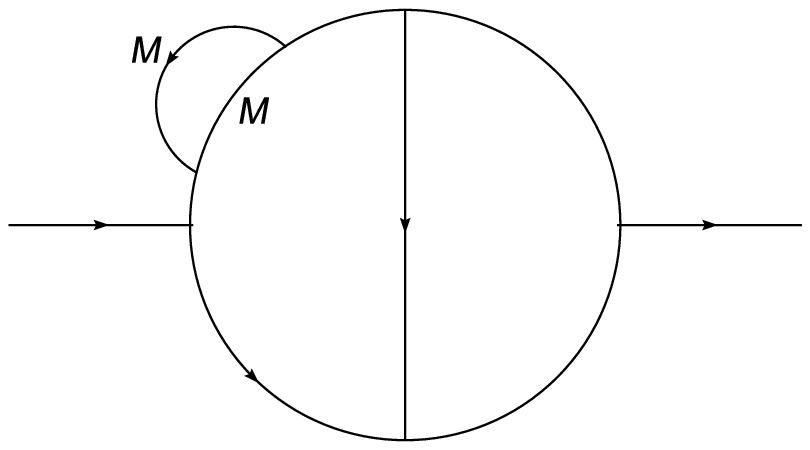}
\end{minipage},etc.
\end{equation*}%
We have shown that IBP+IBFE is a technique that may extend the classes of
diagrams that can be evaluated through this simple procedure, beyond the
scenario given in \cite{IGoIBFE}\ and, although here we have not considered
the IBP in his most general version, we assume that the combination IBP+IBFE
can be also useful beyond the triangle identity.

\bigskip

\bigskip

\textbf{Acknowledgements :}

We acknowledge support from Fondecyt under Grant No. 3080029 and
Grant No.
1095217 and also support from the Centro de Estudios Subatomicos, Valpara%
\'{\i}so, Chile.\newpage

\appendix{\Huge Appendix}

In what follows, we will prove that the solutions that we obtained for the
one-loop example given in Eq. $\left( \ref{ibfe16}\right) $: the first one
using only IBFE $\left( \ref{ibfe20}\right) $ and the second one through the
combined technique IBP-IBFE $\left( \ref{ibfe19}\right) $ are equivalent.
Basically, we have to show that the functions defined below, $G_{\alpha }$
and $G_{\beta }$, are equal:

\begin{equation}
\begin{array}{ll}
G_{\alpha }= & \dfrac{\Gamma \left( 1-\epsilon \right) }{\Gamma \left(
2-\epsilon \right) }\left[ \;_{3}F_{2}\left( \left.
\begin{array}{c}
\begin{array}{ccccc}
2 & , & 1+\epsilon & , & 1-\epsilon%
\end{array}
\\
\begin{array}{ccc}
2-\epsilon & , & 2-\epsilon%
\end{array}%
\end{array}%
\right\vert x\right) \right. \\
&  \\
& -\epsilon \left. \;_{3}F_{2}\left( \left.
\begin{array}{c}
\begin{array}{ccccc}
1 & , & 1+\epsilon & , & 1-\epsilon%
\end{array}
\\
\begin{array}{ccc}
2-\epsilon & , & 2-\epsilon%
\end{array}%
\end{array}%
\right\vert x\right) \right]%
\end{array}
\label{ibfe27}
\end{equation}%
and

\begin{equation}
G_{\beta }=\;_{2}F_{1}\left( \left.
\begin{array}{c}
\begin{array}{ccc}
1+\epsilon & , & 1%
\end{array}
\\
\begin{array}{c}
2-\epsilon%
\end{array}%
\end{array}%
\right\vert x\right) .
\end{equation}%
The variable $x$ represents the quotient $\left( \dfrac{s}{m^{2}}\right) $.
In order to prove the equivalence of both solutions we need to remind the
series expression for an hypergeometric function:

\begin{equation}
\;_{q}F_{q-1}\left( \left.
\begin{array}{c}
\left\{ a\right\} \\
\left\{ b\right\}%
\end{array}%
\right\vert x\right) =\sum\limits_{k=0}^{\infty }\frac{\left( a_{1}\right)
_{k}...\left( a_{q}\right) _{k}}{\left( b_{1}\right) _{k}...\left(
b_{q-1}\right) _{k}}\frac{x^{k}}{k!},
\end{equation}%
then we can rewrite $\left( \ref{ibfe27}\right) $\ as a series, as follows :%
\begin{equation}
\begin{array}{ll}
G_{\alpha } & =\dfrac{\Gamma \left( 1-\epsilon \right) }{\Gamma \left(
2-\epsilon \right) }\left[ \sum\limits_{k=0}^{\infty }\dfrac{\left( 2\right)
_{k}\left( 1+\epsilon \right) _{k}\left( 1-\epsilon \right) _{k}}{\left(
2-\epsilon \right) _{k}\left( 2-\epsilon \right) _{k}}\dfrac{x^{k}}{k!}%
-\epsilon \sum\limits_{k=0}^{\infty }\dfrac{\left( 1\right) _{k}\left(
1+\epsilon \right) _{k}\left( 1-\epsilon \right) _{k}}{\left( 2-\epsilon
\right) _{k}\left( 2-\epsilon \right) _{k}}\dfrac{x^{k}}{k!}\right] \\
&  \\
& =\dfrac{\Gamma \left( 1-\epsilon \right) }{\Gamma \left( 2-\epsilon
\right) }\left[ \sum\limits_{k=0}^{\infty }\dfrac{\left( 1+\epsilon \right)
_{k}\left( 1-\epsilon \right) _{k}}{\left( 2-\epsilon \right) _{k}\left(
2-\epsilon \right) _{k}}\left[ \left( 2\right) _{k}-\epsilon \left( 1\right)
_{k}\right] \dfrac{x^{k}}{k!}\right] .%
\end{array}
\label{ibfe29}
\end{equation}%
On the other side, we have that

\begin{equation}
\left( 2\right) _{k}=\frac{\Gamma \left( 2+k\right) }{\Gamma \left( 2\right)
}=\left( 1+k\right) \Gamma \left( 1+k\right) =\left( 1+k\right) \left(
1\right) _{k},
\end{equation}%
which implies that $\left( \ref{ibfe29}\right) $ reduces to the following
equation:

\begin{equation}
G_{\alpha }=\frac{\Gamma \left( 1-\epsilon \right) }{\Gamma \left(
2-\epsilon \right) }\left[ \sum\limits_{k=0}^{\infty }\frac{\left( 1\right)
_{k}\left( 1+\epsilon \right) _{k}\left( 1-\epsilon \right) _{k}}{\left(
2-\epsilon \right) _{k}\left( 2-\epsilon \right) _{k}}\left[ 1+k-\epsilon %
\right] \frac{x^{k}}{k!}\right] ,  \label{ibfe28}
\end{equation}%
Then, we can transform the expression $\left( 1+k-\epsilon \right) $ as
follows:

\begin{equation*}
\left( 1+k-\epsilon \right) =\frac{\Gamma \left( 2+k-\epsilon \right) }{%
\Gamma \left( 1+k-\epsilon \right) }=\frac{\Gamma \left( 2-\epsilon \right)
\left( 2-\epsilon \right) _{k}}{\Gamma \left( 1-\epsilon \right) \left(
1-\epsilon \right) _{k}}.
\end{equation*}%
By replacing this result in $\left( \ref{ibfe28}\right) $\ some algebra
shows finally that

\begin{equation}
G_{\alpha }=\left[ \sum\limits_{k=0}^{\infty }\frac{\left( 1\right)
_{k}\left( 1+\epsilon \right) _{k}}{\left( 2-\epsilon \right) _{k}}\frac{%
x^{k}}{k!}\right] =\;_{2}F_{1}\left( \left.
\begin{array}{c}
\begin{array}{ccc}
1+\epsilon & , & 1%
\end{array}
\\
\begin{array}{c}
2-\epsilon%
\end{array}%
\end{array}%
\right\vert x\right) =G_{\beta }.
\end{equation}%
Although the equivalence has been shown for the kinematical region $%
\left\vert \dfrac{s}{m^{2}}\right\vert <1$, it is also valid for the region $%
\left\vert \dfrac{s}{m^{2}}\right\vert >1$, since both regions are related
through an analytical continuation.

\end{document}